\pdfoutput=1
\documentclass[twocolumn,superscriptaddress,aps,preprintnumbers,amsmath,amssymb,prl,footinbib]{revtex4-1}
\usepackage{graphicx}
\usepackage{bm}
\usepackage{color}
\usepackage{amsmath,amssymb}	
\usepackage[colorlinks=true, linkcolor=blue, citecolor=blue,urlcolor=black]{hyperref} 
\usepackage{setspace}
\usepackage{longtable}
\usepackage{booktabs}
\usepackage{comment}
\usepackage{array}
\usepackage{cancel}
\usepackage{enumitem}
\usepackage{physics}

\usepackage{todonotes}

\begin{document}

\def\ps{\mathbf{p}}
\def\PS{\mathbf{P}}

\def\simgt{\mathrel{\lower2.5pt\vbox{\lineskip=0pt\baselineskip=0pt
           \hbox{$>$}\hbox{$\sim$}}}}
\def\simlt{\mathrel{\lower2.5pt\vbox{\lineskip=0pt\baselineskip=0pt
           \hbox{$<$}\hbox{$\sim$}}}}
\def\simprop{\mathrel{\lower3.0pt\vbox{\lineskip=1.0pt\baselineskip=0pt
             \hbox{$\propto$}\hbox{$\sim$}}}}
\def\tr{\mathop{\rm tr}}
\def\SU{\mathop{\rm SU}}

\preprint{IPMU20-0085}

\title{Probing Dark Matter Self-interaction with Ultra-faint Dwarf Galaxies}

\author{Kohei Hayashi}
\affiliation{Astronomical Institute, Tohoku University, Sendai, Miyagi, 980-8578, Japan}

\author{Masahiro Ibe}
\affiliation{ICRR, The University of Tokyo, Kashiwa, Chiba 277-8582, Japan}
\affiliation{Kavli IPMU (WPI), UTIAS, The University of Tokyo, Kashiwa, Chiba 277-8583, Japan}

\author{Shin Kobayashi}
\affiliation{ICRR, The University of Tokyo, Kashiwa, Chiba 277-8582, Japan}

\author{Yuhei Nakayama}
\affiliation{ICRR, The University of Tokyo, Kashiwa, Chiba 277-8582, Japan}

\author{Satoshi Shirai}
\affiliation{Kavli IPMU (WPI), UTIAS, The University of Tokyo, Kashiwa, Chiba 277-8583, Japan}

\date{\today}

\begin{abstract}
Self-interacting dark matter (SIDM) has gathered growing attention as a solution to the small scale problems of the collisionless cold dark matter (DM).
We investigate the SIDM using stellar kinematics of 23 ultra-faint dwarf (UFD) galaxies with the phenomenological SIDM halo model. The UFDs are DM-dominated and have less active star formation history.  Accordingly, they are the ideal objects to test the SIDM, as their halo profiles are least affected by the baryonic feedback processes. We found no UFDs favor non-zero self-interaction and some provide stringent constraints within the simple SIDM modeling.
Our result challenges the simple modeling of the SIDM, which urges further investigation of the subhalo dynamical evolution of the SIDM.
\end{abstract}

\maketitle

\section{Introduction}
Collisionless cold dark matter (CDM) has successfully explained the large scale structures of the Universe, such as the galaxy clusters (e.g.~\cite{Vogelsberger:2014kha,Vogelsberger:2014dza,Schaye:2014tpa,Springel:2017tpz}). On smaller scales than the galaxies, however, it has been pointed out that some observed features seem to conflict with the CDM predictions (for reviews, see Refs.\,\cite{Tulin:2017ara,Bullock:2017xww}). For example, observations of the rotation curves of the gas-rich dwarf galaxies favor cored dark matter (DM) halo profiles, while the CDM predicts cuspy profiles~\cite{McGaugh_1998,de_Blok_2001,Oh_2011,Read:2017lvq}.

Self-interacting dark matter (SIDM)~\cite{Spergel_2000} has gathered growing attention as a solution to those discrepancies (see also Ref.~\cite{Tulin:2017ara} for a review). The DM scattering thermalizes the inner halo and makes the isothermal cores.
The SIDM provides a consistent fit to the DM halo profile of the dwarf galaxies for a scattering cross sections per unit mass $\sigma/m$ of $\order{1}$\,cm$^2$/g~\cite{Vogelsberger:2012ku,Rocha:2012jg,Peter:2012jh,Zavala:2012us,Elbert:2014bma,Fry:2015rta}%
~\footnote{For high surface brightness galaxies, the inner halo may be cuspy in the SIDM model due to the gravitational potential from baryons~\cite{Kaplinghat_2014,Vogelsberger_2014,Kaplinghat_2016,Kamada_2017,Creasey_2017,Elbert_2018,Robertson_2018,Ren_2019}. This behavior allows the SIDM model to solve the diversity problem~\cite{Oman_2015,Oman_2016}.}.  
The required cross-section, however, has a tension with the upper limits, $\sigma/m \lesssim 0.1$--$1$\,cm$^2$/g, obtained from the studies of merging and relaxed galaxy clusters~\cite{Markevitch_2004,Randall_2008,Kahlhoefer_2013,Harvey_2015,Robertson_2016,Wittman_2018,Harvey_2019,Bondarenko:2020mpf,Sagunski:2020spe}. 
Due to this tension, more interests are put on the velocity-dependent self-interaction, which evade the cluster constraints.

While the SIDM provides a consistent fit to the dwarf galaxies, however, it is still not conclusive whether the collisionless CDM cannot explain those features. The discrepancies between the observations and the CDM predictions are based on the  simulations without the baryonic effects. The baryonic effects could alter the predictions and make the collisionless CDM consistent with the observations. For example, bursty star formation might cause CDM to heat up at the center of the dwarf galaxies and form the cored DM 
profile~\cite{Read:2004xc,Mashchenko:2007jp,Pontzen:2011ty,Onorbe:2015ija,Tollet:2015gqa,Hayashi:2020jze}. To separate those possibilities, it is ideal to investigate the DM self-interaction in the environments with fewer baryons.

In this letter, we discuss the constraints on the DM self-interaction cross-section by comparing the phenomenological modeling of the SIDM halo profile~\cite{Kaplinghat_2016,Valli:2017ktb} and the stellar kinematics of the 23 ultra-faint dwarf (UFD) galaxies. The UFDs are considered to be more DM-dominated and have less active star formation history. Besides, the typical DM velocities of the UFDs, 
$v=\order{10}\,\mathrm{km/s}$, are close to those of the dwarf galaxies which seem to favor the SIDM. Therefore, the UFDs are ideal environments to derive robust constraints on the velocity-dependent cross-section at the typical velocity of $v\sim 10$ km/s.

\section{Models}
The stellar motion in a DM dominated system such as a dwarf spheroidal galaxy~(dSph) is governed by the DM gravitational potential, $\Phi_{\mathrm{DM}}$.
Assuming that an UFD is a spherically symmetric and DM dominated steady-state system, the DM potential relates to the moments of the stellar distribution function through the Jeans equation~\cite{2008gady.book.....B}:
\begin{align}
\label{eq:Jeans}
   \frac{\partial\nu(r) \sigma_r^2(r)}{\partial r} 
        + 
    \frac{2\beta_{\mathrm{ani}}(r)\sigma_r^2(r)}{r} = - \nu(r) \frac{\partial \Phi_{\mathrm{DM}}}{\partial r},
\end{align}
where $r$ denotes the radius from the center of the UFD and $\nu(r)$ is the intrinsic stellar distribution. 
The velocity dispersions of the stars are defined by $\sigma_r, \sigma_{\theta}$ and $\sigma_{\phi}$ in a spherical coordinate system.
Here, from spherical symmetry, we take $\sigma_{\theta}$ = $\sigma_{\phi}$, and the stellar orbital anisotropy is defined by $\beta_{\mathrm{ani}}(r)\equiv 1-\sigma^2_{\theta}/\sigma^2_r$. 
In this work, we adopt a general and realistic anistropy modeling derived by Ref.~\cite{Baes:2007tx}:
\begin{align}
\label{eq:anisotropy}
\beta_{\mathrm{ani}}(r) = \frac{\beta_0+\beta_{\infty}(r/r_{\beta})^{\eta}}{1+(r/r_{\beta})^{\eta}},
\end{align}
where the four parameters are the inner anisotropy $\beta_0$, the outer anisotropy $\beta_{\infty}$, the sharpness of the transition $\eta$, and the transition radius $r_{\beta}$.

Solving Eq.\,\eqref{eq:Jeans}, we obtain the intrinsic velocity dispersion profiles, $\sigma_r(r)$.
However, such intrinsic dispersions are not directly observed, and only the line-of-sight velocity distribution and the stellar surface density profile are measurable.
Therefore, we integrate $\sigma_r(r)$ along the line-of-sight  \cite{2008gady.book.....B},
\begin{align}
\label{eq:losdisp}
\sigma_{\rm l.o.s.}^2(R)=\frac{2}{\Sigma(R)}\int_R^\infty\!\! dr \Bigl(1-\beta_{\mathrm{ani}}(r)\frac{R^2}{r^2}\Bigr)\frac{\nu(r)\sigma_r^2(r)}{\sqrt{1-R^2/r^2}},
\end{align}
where $R$ is a projected radius, and $\Sigma(R)$ is the surface density profile derived from the intrinsic stellar density $\nu(r)$ through an Abel transform.
For the stellar density profile, we adopt the Plummer profile~\cite{Plummer:1911zza}.

In order to determine the gravitational potential of the SIDM halo, we adopt the halo model 
in Refs.\,\cite{Kaplinghat_2016,Valli:2017ktb}.
In the vicinity of the center of the DM halo, we assume that the halo profile can be described by isothermal halo model, 
\begin{align}
    h''(x)+ 2 \frac{h'(x)}{x} = -e^{h(x)},~\lim_{x\to 0}h(x) = \lim_{x\to 0}h'(x) = 0.
\end{align}
Here, we define a normalized halo density, $h\equiv \ln(\rho/\rho_0)$ and a normalized radius $x=r/r_c$ ($r_c = \sigma_0/\sqrt{4\pi G\rho_0}$), where $\rho_0$ and $\sigma_0$ denote the DM density and the velocity dispersion at the center of the halo\,\footnote{
Here, the mass density and the pressure of DM are related by $p_{DM}=\sigma_0^2\rho_{DM}$.
This corresponds to the Maxwell-Boltzmann velocity distribution, $f_{DM} \propto e^{-v^2/2\sigma_0^2}$.
We do not truncate the distribution by the escape velocity.
In this case, the mean relative DM velocity is given by, $v = 4\sigma_0/\sqrt{\pi}$.
}, respectively.
On the other hand, at a larger scale than some radius $r>r_1$, we assume that the halo is described by the Navarro-Frenk-White (NFW) profile~\cite{Navarro:1996gj},
\begin{align}
    \rho = \frac{\rho_s}{r/r_s(1+r/r_s)^2}.
\end{align}
At the halo transition radius $r=r_1$, we impose
continuity conditions of
the density and the mass,
\begin{align}
\label{eq:continuity}
    \rho_0 e^{h(x_{c1})} = \frac{\rho_s}{x_{s1}(1+x_{s1})^2},~
    \mathcal{R}_{\mathrm{iso}}(r_1) = \mathcal{R}_{\mathrm{NFW}}(r_1),
\end{align}
where $\mathcal{R} = M(r)/(4\pi r^3\rho(r))$ is the ratio of the mass and the density, $M(r) = \int_0^r d^3r'\rho(r')$ is the total halo mass within the radius $r$, $x_{c1}\equiv r_1/r_c$ and $x_{s1}\equiv r_1/r_s$.
As the central isothermal region is formed by the DM self-interaction, $r_1$
is determined by the self-interaction cross section via,
\begin{align}
\label{eq:r1 from scattering}
    \rho(r_1)
    \frac{\langle\sigma v\rangle}{m}\simeq \rho(r_1)
    \frac{\sigma \langle v\rangle}{m}\simeq\frac{\sqrt{\pi}}{4t_{\mathrm{age}}}.
\end{align}
Here, $\sigma$ is the velocity averaged scattering cross section and we have used the relation between the mean relative velocity, $\langle v\rangle$, and $\sigma_0$, $\langle v\rangle = 4\sigma_0/\sqrt{\pi}$.
With Eqs.\,\eqref{eq:continuity} and \eqref{eq:r1 from scattering}, we derive the isothermal profile parameters $(\rho_0,\sigma_0)$ when the NFW profile parameters $(\rho_s, r_s)$,  $\sigma/m$, and $t_{\mathrm{age}}$ are given.

When small values of $(\rho_s,r_s,\sigma/m\times t_{\mathrm{age}})$ are given, there is a unique solution for Eqs.\,\eqref{eq:continuity} and \eqref{eq:r1 from scattering}.
However, there can be multiple solutions for relatively large values of $(\rho_s,r_s,\sigma/m\times t_{\mathrm{age}})$.
In that case, we choose the solution which has the minimum value of $r_1$ among the multiple solutions.
This is because
we assume that the SIDM halo profile gradually evolves from the NFW profile and the isothermal core, i.e., $r_1$, is smaller in the past.
Note that for parameters satisfying
\begin{align}
    \frac{\sigma}{m} \gtrsim 9~{\rm cm^2/g} \! \left( \frac{ 0.1 M_{\odot}/{\rm pc}^{3}}{\rho_{s}}\right )^{\!3/2} \!\!\!
   \left( \frac{1\,{\rm kpc}}{r_{s}}\right )\!\!
   \left( \frac{10\,{\rm Gyr}}{t_{\mathrm{age}}}\right )\!,
\end{align}
there is no solution consistent with continuity conditions  \eqref{eq:continuity} and \eqref{eq:r1 from scattering}.
For such a large cross-section, the description with the isothermal core + NFW outskirt is no more valid, and hence, we discard such a large cross-section in this analysis.

In Ref.\,\cite{Valli:2017ktb},
the 8 classical dSphs are used to investigate the DM self-interacting cross-section. 
There, it is required that the parameters of the NFW profiles fall in the region which encompasses those of the most-massive 15 subhalos from the CDM-only simulation of Ref.\,\cite{Vogelsberger:2015gpr}.
This requirement is motivated to study how the SIDM ameliorates the too-big-to-fail problem
of the underlying CDM-only simulaiton~\cite{BoylanKolchin:2011de,BoylanKolchin:2011dk}.
In our analysis, on the other hand, we do not employ this requirement since we use the UFDs, which are not in conflict with this problem.
Instead, we require that the NFW parameters satisfy the concentration-mass relation of the subhalos suggested by the simulations in Ref.\,\cite{Moline:2016pbm} (see also Ref.\,\cite{Ishiyama:2019hmh}):
\begin{eqnarray}
c^0_{200}(M_{200},x_{\mathrm{sub}}) = && c_0 \left[1+\sum^{3}_{i=1} \left[a_i \log_{10}\left(\frac{M_{200}}{10^8 h^{-1}M_{\odot}}\right)\right]^i \right] \nonumber \\
&& \times \left[1+b\log_{10}(x_{\mathrm{sub}})\right]. \label{eq:c200}
\end{eqnarray}
Here, $c_0=19.9$, $a_i=\{-0.195,0.089,0.089\}$, and $b=-0.54$ which are the best-fit parameters for the concentration-mass relation.
$M_{200}$ is the enclosed mass within $r_{200}$ in which the spherical overdensity is 200 times the critical density of the Universe, $\rho_{\mathrm{crit}}$, that is $M_{200}=200 \times (4\pi/3)r^3_{200}\rho_{\mathrm{crit}}$.
$x_{\mathrm{sub}}\equiv r_{\mathrm{sub}}/r_{200,\mathrm{host}}$ is a subhalo distance from the center of a host halo divided by $r_{200}$ of the host halo.
This requirement is based on the modeling of Refs.\,\cite{Kaplinghat_2016,Valli:2017ktb} in which the NFW parameters should be the ones not affected by the DM self-interaction.

\begin{figure*}[t!]
    \centering
    \includegraphics[width=\textwidth]{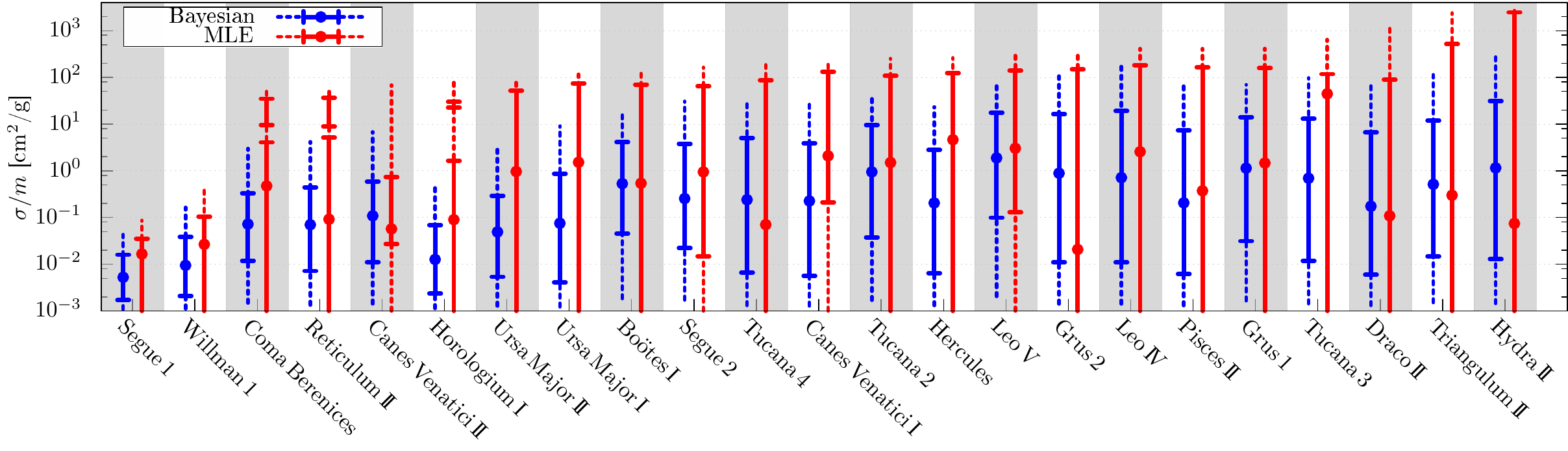}

    \caption{
    The interval estimates of $\sigma/m$ for the 23 UFDs.
    The solid (dotted) segments show $1\sigma~(2\sigma)$ intervals.
    The blue segments show the Bayesian analysis with the log-flat prior of $\sigma/m$.
    The red segments show the credible intervals of $\sigma/m$ with MLE.
    }
    \label{fig:result}
\end{figure*}

In order to obtain $x_{\mathrm{sub}}$, we need to estimate $r_{200}$ of the Milky Way halo, $r_{200,\mathrm{MW}}$.
The masses of the Milky Way estimated from a variety of methods are consistent with $M_{200,\mathrm{MW} }\simeq1\times10^{12}M_{\odot}$, but there exists a large uncertainty with $M_{200,\mathrm{MW}}\simeq0.5-2.0\times10^{12}M_{\odot}$~\cite{Wang:2019ubx}.
Therefore, $r_{200,\mathrm{MW}}$ associated with the mass errors also has a large uncertainty, $r_{200,\mathrm{MW}}\simeq 210\pm50$~kpc.
From Eq.\,\eqref{eq:c200}, the error of $r_{200,\mathrm{MW}}$, however, may give only a small impact on the concentration. 
Thus, we fix $r_{200,\mathrm{MW}}=210$~kpc.

\section{Data}
In this work, we investigate SIDM properties for 23 UFDs associated with the Milky Way (Segue~1, Segue~2, Bo\"otes~I, Hercules, Coma~Bernices, Canes~Venatici~I, Canes~Venatici~II, Leo~IV, Leo~V, Ursa~Major~I, Ursa~Major~II, Reticulum~II, Draco~II, Triangulum~II, Hydra~II, Pisces~II, Grus~1, Grus~2, Horologium~I, Tucana~2, Tucana~3, Tucana~4, and Willman~1).

The basic structural properties (the positions of the centers, the distances, and the half-light radii with the Plummer profile) of their galaxies are adopted from the original observation papers~\cite{2008ApJ...684.1075M,2009MNRAS.397.1748B,2015ApJ...805..130K,2015ApJ...804L...5M,2012ApJ...756...79S,2015ApJ...813..109D,2015ApJ...807...50B,2015ApJ...805..130K,2018ApJ...860...66M}.
For the stellar-kinematics of their member stars, we utilize the currently available data taken from each spectroscopic observation paper~\cite{2011ApJ...733...46S,2013ApJ...770...16K,2011ApJ...736..146K,2007ApJ...670..313S, 2015ApJ...808...95S, 2016MNRAS.458L..59M,2017ApJ...838...83K,2015ApJ...810...56K,2016ApJ...819...53W,2020ApJ...892..137S,2015ApJ...811...62K,2017ApJ...838...11S,2020ApJ...892..137S,2011AJ....142..128W}.
The membership selections for each galaxy follow the methods described in the cited papers.
The unresolved binary stars in a stellar kinematic sample may affect the measured velocity dispersion of our target galaxies due to binary orbital motion.
However, several papers show that binary star candidates can be excluded from the member stars and suggest that such an effect is much smaller than the measurement uncertainty of the velocity.
Therefore, we suppose that the effect of binaries can be negligible.

\section{Analysis and Results}
We perform the fitting with the likelihood function $\log ({\cal L}_{\rm tot}) = \log({\cal L}_{\rm dis})+ \log({\cal L}_{\rm CDM}) +\log({\cal L}_{r_{1/2}})$.
The contribution from the stellar kinematic data is calculated by,
\begin{align}
    -2 \log({\cal L}_{\rm dis}) = \sum_{i} \left[ \frac{(v_i -V)^2}{\sigma_{i}^2} + \log(2\pi \sigma_i^2) \right]. \label{eq:like_dis}
\end{align}
Here, $i$ runs the member stars of each UFD with the line-of-sight velocity $v_i$ and $V$ is the mean line-of-sight velocity of the member stars.
The dispersion of the line-of-sight velocity $\sigma_{i}^2$ is the squared sum of the intrinsic dispersion in Eq.\,\eqref{eq:losdisp} and the measurement error $\varepsilon_i$: $\sigma_i^2 = \sigma_{\rm l.o.s.}^2(R_i) + \varepsilon_i^2$.
We always take $V$ to maximize the likelihood ${\cal L}_{\rm dis}$, i.e., $d \log({\cal L}_{\rm dis})/dV = 0$.

Besides, we impose the concentration-mass relation for the NFW parameters, by including the likelihood, 
\begin{align}
    -2 \log({\cal L}_{\rm CDM}) = \frac{(\log_{10}(c_{200}) - \log_{10}(c^0_{200}))^2}{\sigma^2_{\rm CDM}}.
\end{align}
Here, $c_{200}$ is estimated from the NFW parameters and $c^{0}_{200}$ is the median subhalo concentration-mass relation in Eq.\,\eqref{eq:c200}
with $\sigma_{\rm CDM} = 0.13$~\cite{Moline:2016pbm}.

We also add the uncertainty of the half-light radius $r_{1/2}$ in the Plummer profile by adding $-2\log({\cal L}_{r_{1/2}}) = (r_{1/2} - r^{0}_{1/2} )^2/\delta r_{1/2}^2 $, where $r^{0}_{1/2}$ and $\delta r_{1/2}$ are the measured value and its error, respectively.

\begin{figure}[htbp]
    \centering
    \includegraphics[width=0.5\textwidth]{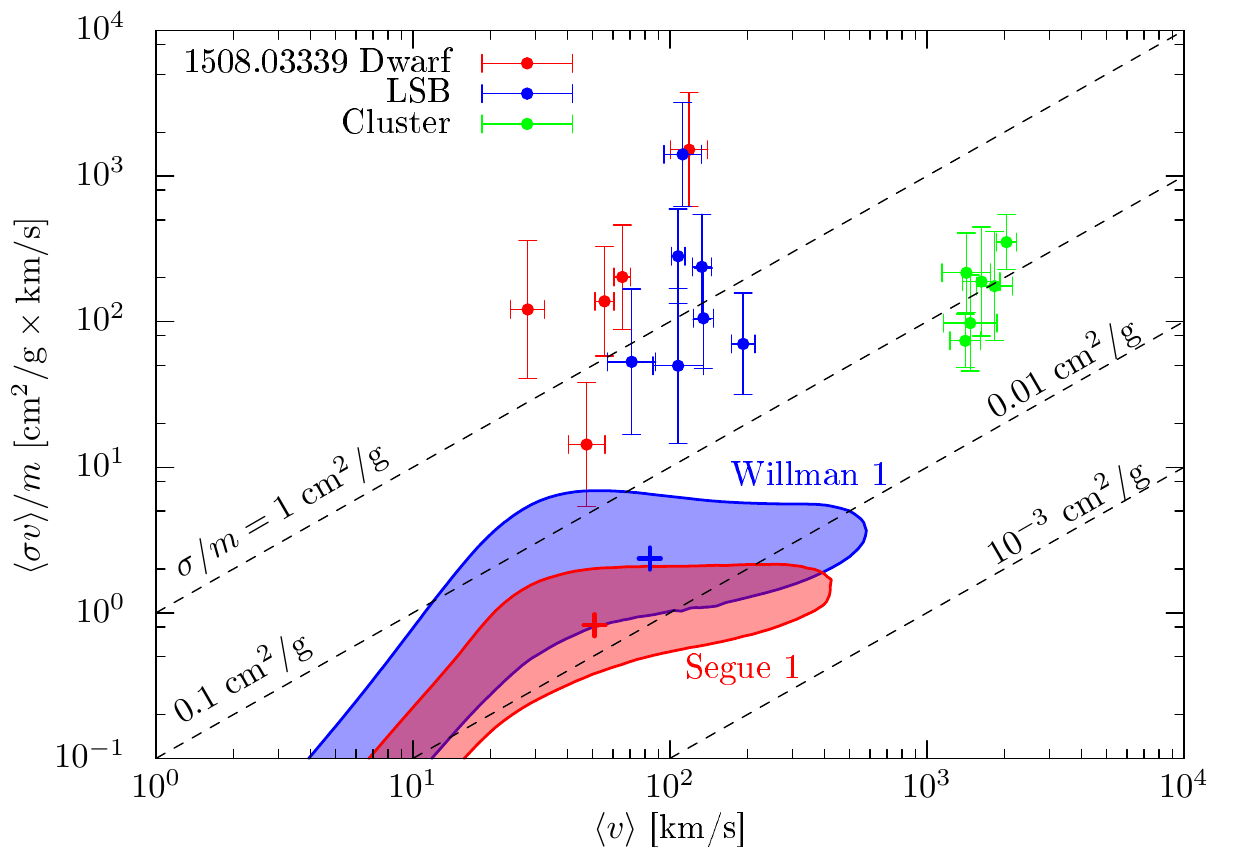}
    \caption{The $1\sigma$ parameter estimation of $\langle v \rangle \mbox{-}\langle \sigma v \rangle/m $ based on MLE. 
    We also show the SIDM cross-section which are favored by the dwarf irregular galaxies (red), low surface brightness galaxies (blue) and clusters (green).}
    \label{fig:sigmav}
\end{figure}

Using the likelihood function, $\log ({\cal L}_{\rm tot})$, 
we fit the $8$ parameters,
$\rho_s, r_s$, 
$\beta_0, \beta_\infty, r_\beta, \eta$,
$r_{1/2}$, and $\sigma/m$.
In the Bayesian analysis,
we assume that the prior distributions of 
the NFW parameters and the half-light radius are 
the flat distributions of the following expressions in
the ranges of
$-4 \le \log_{10}(\rho_s~[M_{\odot}/{\rm pc}^3]) \le 2,~0 \le \log_{10}(r_{s, \beta}~[{\rm pc}]) \le 4$, $1 \le \eta \le 10, ~0 \le 2^{\beta_{0(\infty)}} \le 1(2)$ and $0 \le r_{1/2}~[{\rm pc}] \le 1000$, respectively.
As for the SIDM cross-section, we consider
the log-flat distribution in $-3 \le \log_{10}(\sigma/m~[{\rm cm^2/g}]) \le 3 $.
We estimate the posterior distribution via Markov chain Monte Carlo by the algorithm of Ref.~\cite{2010CAMCS...5...65G}.
We also perform the fitting via maximum likelihood estimation (MLE), which provides prior-independent constraints.

In Fig.\,\ref{fig:result}, we show the fitting results of $\sigma/m$ for  $t_\mathrm{age} = 10\,\mathrm{Gyr}$.
For the Bayesian analysis, we show the median of $\sigma/m$ with $1\sigma$ and $2\sigma$ credible intervals.
The figure shows that the posterior distributions of $\sigma/m$ reach the lower limit of its prior distribution.
Thus, no UFDs strongly favor non-zero self-interaction cross-section.
This also means that the upper limit of $\sigma/m$ strongly depends on the prior distribution
in the Bayesian analysis.

For the MLE, we estimate the $1(2)\sigma$ confidence intervals via $-2 \Delta  \log({\cal L}_{\rm tot})=1$ and $4$.
We have checked that the estimated intervals by the MLE are consistent with the bootstrap Monte-Carlo simulation.
The figure again shows that no UFDs favor non-zero cross-section.
In particular, the fitting results for Segue~1 and Willman~1 are consistent with zero cross-section at $1\sigma$ C.L. and provide the $2\sigma$ upper-limit $8.6\times 10^{-2}~{\rm cm^2/g}$ and  $0.39~{\rm cm^2/g}$, respectively.

In Fig.\,\ref{fig:sigmav}, we show the $1\sigma$ estimation of $\langle v \rangle \mbox{-} \langle \sigma v \rangle/m $ with the MLE for Segue~1 and Willman~1, which correspond to two dimensional contours of $-2 \Delta  \log({\cal L}_{\rm tot})=2.3$.
Compared with the favored SIDM cross-section in the previous study using the dwarf irregular galaxies~\cite{Oh_2011}, the low surface brightness galaxies~\cite{de_Naray_2008} (blue), and galaxy clusters~\cite{Newman_2013} (green) (see Ref.~\cite{Kaplinghat_2016} for details),
the Segue~1 and Willman~1 place stringent upper limits.

\section{Conclusion}
We investigated the SIDM by using the stellar kinematics of the 23 UFD galaxies with the phenomenological modeling of the SIDM halo profile.
We found all the UFD galaxies are consistent with collisionless DM.
In particular, Segue~1 and Willman~1 provide stringent constraints on the self-interacting cross-section: $\sigma/m < \order{0.1}~\mathrm{cm^2/g}$.
As seen in Fig.\,\ref{fig:sigmav}, our result with the UFDs is in considerable tension with the DM self-interaction strength preferred by the dwarf irregular and the low surface brightness galaxies which are more affected by the baryonic feedback effects than the UFDs.
In the present framework, it would not be easy to explain this discrepancy with the DM velocity dependent cross-section, as the typical velocities of the DM in the UFDs and those in the dwarf irregular galaxies are close with each other
\footnote{For a model which predicts 
the SIDM cross-section with an intricate velocity dependence, see e.g. Ref.\cite{Chu:2018fzy}.}.

It should be emphasized that our analysis is based on the simple steady-state modeling of the SIDM in Refs.~\cite{Kaplinghat_2016,Valli:2017ktb}.
If some of the UFDs are in the core-collapse phase,
for example, the simple model does not describe 
their halo profiles properly, 
which could invalidate 
the constraints obtained in this work.
(See Ref.~\cite{Correa:2020qam} for the study of the SIDM using the classical dSphs in which the core-collapse process is taken into account semi-analytically~\cite{Balberg:2002ue}.) 
The estimation of the core-collapse time scale of the subhalo, however, may
depend on subtle dynamics such as the tidal-stripping, and hence, needs further studies~\cite{Vogelsberger:2012ku,Rocha:2012jg,Elbert:2014bma,Nishikawa:2019lsc,Kummer:2019yrb,Robles:2019mfq,Sameie:2019zfo,Kahlhoefer:2019oyt}.

\begin{acknowledgments}

We thank H. Yu, M. Valli and S. Ando for helpful discussions.
This work is supported by Grant-in-Aid for Scientific Research from the Ministry of Education, Culture, Sports, Science, and Technology (MEXT), Japan, 17H02878 (M.I. and S.S.), 18H05542 (M.I.) 18K13535, 19H04609, 20H01895 (S.S.) and 20H01895 (K.H.), and by World Premier International Research Center Initiative (WPI), MEXT, Japan. 
This work is also supported by the Advanced Leading Graduate Course for Photon Science (S.K.), the JSPS Research Fellowships for Young Scientists (S.K.) and International Graduate Program for Excellence in Earth-Space Science (Y.N.).
\end{acknowledgments}

%%%%%%%%%%%%% References %%%%%%%%%%%%%%%%%%%
\bibliographystyle{apsrev4-1}
\bibliography{ref}

\begin{thebibliography}{88}%
\makeatletter
\providecommand \@ifxundefined [1]{%
 \@ifx{#1\undefined}
}%
\providecommand \@ifnum [1]{%
 \ifnum #1\expandafter \@firstoftwo
 \else \expandafter \@secondoftwo
 \fi
}%
\providecommand \@ifx [1]{%
 \ifx #1\expandafter \@firstoftwo
 \else \expandafter \@secondoftwo
 \fi
}%
\providecommand \natexlab [1]{#1}%
\providecommand \enquote  [1]{``#1''}%
\providecommand \bibnamefont  [1]{#1}%
\providecommand \bibfnamefont [1]{#1}%
\providecommand \citenamefont [1]{#1}%
\providecommand \href@noop [0]{\@secondoftwo}%
\providecommand \href [0]{\begingroup \@sanitize@url \@href}%
\providecommand \@href[1]{\@@startlink{#1}\@@href}%
\providecommand \@@href[1]{\endgroup#1\@@endlink}%
\providecommand \@sanitize@url [0]{\catcode `\\12\catcode `\$12\catcode
  `\&12\catcode `\#12\catcode `\^12\catcode `\_12\catcode `\%12\relax}%
\providecommand \@@startlink[1]{}%
\providecommand \@@endlink[0]{}%
\providecommand \url  [0]{\begingroup\@sanitize@url \@url }%
\providecommand \@url [1]{\endgroup\@href {#1}{\urlprefix }}%
\providecommand \urlprefix  [0]{URL }%
\providecommand \Eprint [0]{\href }%
\providecommand \doibase [0]{http://dx.doi.org/}%
\providecommand \selectlanguage [0]{\@gobble}%
\providecommand \bibinfo  [0]{\@secondoftwo}%
\providecommand \bibfield  [0]{\@secondoftwo}%
\providecommand \translation [1]{[#1]}%
\providecommand \BibitemOpen [0]{}%
\providecommand \bibitemStop [0]{}%
\providecommand \bibitemNoStop [0]{.\EOS\space}%
\providecommand \EOS [0]{\spacefactor3000\relax}%
\providecommand \BibitemShut  [1]{\csname bibitem#1\endcsname}%
\let\auto@bib@innerbib\@empty
%</preamble>
\bibitem [{\citenamefont {Vogelsberger}\ \emph
  {et~al.}(2014{\natexlab{a}})\citenamefont {Vogelsberger}, \citenamefont
  {Genel}, \citenamefont {Springel}, \citenamefont {Torrey}, \citenamefont
  {Sijacki}, \citenamefont {Xu}, \citenamefont {Snyder}, \citenamefont {Bird},
  \citenamefont {Nelson},\ and\ \citenamefont
  {Hernquist}}]{Vogelsberger:2014kha}%
  \BibitemOpen
  \bibfield  {author} {\bibinfo {author} {\bibfnamefont {M.}~\bibnamefont
  {Vogelsberger}}, \bibinfo {author} {\bibfnamefont {S.}~\bibnamefont {Genel}},
  \bibinfo {author} {\bibfnamefont {V.}~\bibnamefont {Springel}}, \bibinfo
  {author} {\bibfnamefont {P.}~\bibnamefont {Torrey}}, \bibinfo {author}
  {\bibfnamefont {D.}~\bibnamefont {Sijacki}}, \bibinfo {author} {\bibfnamefont
  {D.}~\bibnamefont {Xu}}, \bibinfo {author} {\bibfnamefont {G.~F.}\
  \bibnamefont {Snyder}}, \bibinfo {author} {\bibfnamefont {S.}~\bibnamefont
  {Bird}}, \bibinfo {author} {\bibfnamefont {D.}~\bibnamefont {Nelson}}, \ and\
  \bibinfo {author} {\bibfnamefont {L.}~\bibnamefont {Hernquist}},\ }\href
  {\doibase 10.1038/nature13316} {\bibfield  {journal} {\bibinfo  {journal}
  {Nature}\ }\textbf {\bibinfo {volume} {509}},\ \bibinfo {pages} {177}
  (\bibinfo {year} {2014}{\natexlab{a}})},\ \Eprint
  {http://arxiv.org/abs/1405.1418} {arXiv:1405.1418 [astro-ph.CO]} \BibitemShut
  {NoStop}%
\bibitem [{\citenamefont {Vogelsberger}\ \emph
  {et~al.}(2014{\natexlab{b}})\citenamefont {Vogelsberger}, \citenamefont
  {Genel}, \citenamefont {Springel}, \citenamefont {Torrey}, \citenamefont
  {Sijacki}, \citenamefont {Xu}, \citenamefont {Snyder}, \citenamefont
  {Nelson},\ and\ \citenamefont {Hernquist}}]{Vogelsberger:2014dza}%
  \BibitemOpen
  \bibfield  {author} {\bibinfo {author} {\bibfnamefont {M.}~\bibnamefont
  {Vogelsberger}}, \bibinfo {author} {\bibfnamefont {S.}~\bibnamefont {Genel}},
  \bibinfo {author} {\bibfnamefont {V.}~\bibnamefont {Springel}}, \bibinfo
  {author} {\bibfnamefont {P.}~\bibnamefont {Torrey}}, \bibinfo {author}
  {\bibfnamefont {D.}~\bibnamefont {Sijacki}}, \bibinfo {author} {\bibfnamefont
  {D.}~\bibnamefont {Xu}}, \bibinfo {author} {\bibfnamefont {G.~F.}\
  \bibnamefont {Snyder}}, \bibinfo {author} {\bibfnamefont {D.}~\bibnamefont
  {Nelson}}, \ and\ \bibinfo {author} {\bibfnamefont {L.}~\bibnamefont
  {Hernquist}},\ }\href {\doibase 10.1093/mnras/stu1536} {\bibfield  {journal}
  {\bibinfo  {journal} {Mon. Not. Roy. Astron. Soc.}\ }\textbf {\bibinfo
  {volume} {444}},\ \bibinfo {pages} {1518} (\bibinfo {year}
  {2014}{\natexlab{b}})},\ \Eprint {http://arxiv.org/abs/1405.2921}
  {arXiv:1405.2921 [astro-ph.CO]} \BibitemShut {NoStop}%
\bibitem [{\citenamefont {Schaye}\ \emph {et~al.}(2015)\citenamefont {Schaye}
  \emph {et~al.}}]{Schaye:2014tpa}%
  \BibitemOpen
  \bibfield  {author} {\bibinfo {author} {\bibfnamefont {J.}~\bibnamefont
  {Schaye}} \emph {et~al.},\ }\href {\doibase 10.1093/mnras/stu2058} {\bibfield
   {journal} {\bibinfo  {journal} {Mon. Not. Roy. Astron. Soc.}\ }\textbf
  {\bibinfo {volume} {446}},\ \bibinfo {pages} {521} (\bibinfo {year}
  {2015})},\ \Eprint {http://arxiv.org/abs/1407.7040} {arXiv:1407.7040
  [astro-ph.GA]} \BibitemShut {NoStop}%
\bibitem [{\citenamefont {Springel}\ \emph {et~al.}(2018)\citenamefont
  {Springel} \emph {et~al.}}]{Springel:2017tpz}%
  \BibitemOpen
  \bibfield  {author} {\bibinfo {author} {\bibfnamefont {V.}~\bibnamefont
  {Springel}} \emph {et~al.},\ }\href {\doibase 10.1093/mnras/stx3304}
  {\bibfield  {journal} {\bibinfo  {journal} {Mon. Not. Roy. Astron. Soc.}\
  }\textbf {\bibinfo {volume} {475}},\ \bibinfo {pages} {676} (\bibinfo {year}
  {2018})},\ \Eprint {http://arxiv.org/abs/1707.03397} {arXiv:1707.03397
  [astro-ph.GA]} \BibitemShut {NoStop}%
\bibitem [{\citenamefont {Tulin}\ and\ \citenamefont
  {Yu}(2018)}]{Tulin:2017ara}%
  \BibitemOpen
  \bibfield  {author} {\bibinfo {author} {\bibfnamefont {S.}~\bibnamefont
  {Tulin}}\ and\ \bibinfo {author} {\bibfnamefont {H.-B.}\ \bibnamefont {Yu}},\
  }\href {\doibase 10.1016/j.physrep.2017.11.004} {\bibfield  {journal}
  {\bibinfo  {journal} {Phys. Rept.}\ }\textbf {\bibinfo {volume} {730}},\
  \bibinfo {pages} {1} (\bibinfo {year} {2018})},\ \Eprint
  {http://arxiv.org/abs/1705.02358} {arXiv:1705.02358 [hep-ph]} \BibitemShut
  {NoStop}%
\bibitem [{\citenamefont {Bullock}\ and\ \citenamefont
  {Boylan-Kolchin}(2017)}]{Bullock:2017xww}%
  \BibitemOpen
  \bibfield  {author} {\bibinfo {author} {\bibfnamefont {J.~S.}\ \bibnamefont
  {Bullock}}\ and\ \bibinfo {author} {\bibfnamefont {M.}~\bibnamefont
  {Boylan-Kolchin}},\ }\href {\doibase 10.1146/annurev-astro-091916-055313}
  {\bibfield  {journal} {\bibinfo  {journal} {Ann. Rev. Astron. Astrophys.}\
  }\textbf {\bibinfo {volume} {55}},\ \bibinfo {pages} {343} (\bibinfo {year}
  {2017})},\ \Eprint {http://arxiv.org/abs/1707.04256} {arXiv:1707.04256
  [astro-ph.CO]} \BibitemShut {NoStop}%
\bibitem [{\citenamefont {McGaugh}\ and\ \citenamefont
  {de~Blok}(1998)}]{McGaugh_1998}%
  \BibitemOpen
  \bibfield  {author} {\bibinfo {author} {\bibfnamefont {S.~S.}\ \bibnamefont
  {McGaugh}}\ and\ \bibinfo {author} {\bibfnamefont {W.~J.~G.}\ \bibnamefont
  {de~Blok}},\ }\href {\doibase 10.1086/305612} {\bibfield  {journal} {\bibinfo
   {journal} {The Astrophysical Journal}\ }\textbf {\bibinfo {volume} {499}},\
  \bibinfo {pages} {41–65} (\bibinfo {year} {1998})}\BibitemShut {NoStop}%
\bibitem [{\citenamefont {de~Blok}\ \emph {et~al.}(2001)\citenamefont
  {de~Blok}, \citenamefont {McGaugh}, \citenamefont {Bosma},\ and\
  \citenamefont {Rubin}}]{de_Blok_2001}%
  \BibitemOpen
  \bibfield  {author} {\bibinfo {author} {\bibfnamefont {W.~J.~G.}\
  \bibnamefont {de~Blok}}, \bibinfo {author} {\bibfnamefont {S.~S.}\
  \bibnamefont {McGaugh}}, \bibinfo {author} {\bibfnamefont {A.}~\bibnamefont
  {Bosma}}, \ and\ \bibinfo {author} {\bibfnamefont {V.~C.}\ \bibnamefont
  {Rubin}},\ }\href {\doibase 10.1086/320262} {\bibfield  {journal} {\bibinfo
  {journal} {The Astrophysical Journal}\ }\textbf {\bibinfo {volume} {552}},\
  \bibinfo {pages} {L23–L26} (\bibinfo {year} {2001})}\BibitemShut {NoStop}%
\bibitem [{\citenamefont {Oh}\ \emph {et~al.}(2011)\citenamefont {Oh},
  \citenamefont {de~Blok}, \citenamefont {Brinks}, \citenamefont {Walter},\
  and\ \citenamefont {Kennicutt}}]{Oh_2011}%
  \BibitemOpen
  \bibfield  {author} {\bibinfo {author} {\bibfnamefont {S.-H.}\ \bibnamefont
  {Oh}}, \bibinfo {author} {\bibfnamefont {W.~J.~G.}\ \bibnamefont {de~Blok}},
  \bibinfo {author} {\bibfnamefont {E.}~\bibnamefont {Brinks}}, \bibinfo
  {author} {\bibfnamefont {F.}~\bibnamefont {Walter}}, \ and\ \bibinfo {author}
  {\bibfnamefont {R.~C.}\ \bibnamefont {Kennicutt}},\ }\href {\doibase
  10.1088/0004-6256/141/6/193} {\bibfield  {journal} {\bibinfo  {journal} {The
  Astronomical Journal}\ }\textbf {\bibinfo {volume} {141}},\ \bibinfo {pages}
  {193} (\bibinfo {year} {2011})}\BibitemShut {NoStop}%
\bibitem [{\citenamefont {Read}\ \emph {et~al.}(2017)\citenamefont {Read},
  \citenamefont {Iorio}, \citenamefont {Agertz},\ and\ \citenamefont
  {Fraternali}}]{Read:2017lvq}%
  \BibitemOpen
  \bibfield  {author} {\bibinfo {author} {\bibfnamefont {J.}~\bibnamefont
  {Read}}, \bibinfo {author} {\bibfnamefont {G.}~\bibnamefont {Iorio}},
  \bibinfo {author} {\bibfnamefont {O.}~\bibnamefont {Agertz}}, \ and\ \bibinfo
  {author} {\bibfnamefont {F.}~\bibnamefont {Fraternali}},\ }\href {\doibase
  10.1093/mnras/stx147} {\bibfield  {journal} {\bibinfo  {journal} {Mon. Not.
  Roy. Astron. Soc.}\ }\textbf {\bibinfo {volume} {467}},\ \bibinfo {pages}
  {2019} (\bibinfo {year} {2017})},\ \Eprint {http://arxiv.org/abs/1607.03127}
  {arXiv:1607.03127 [astro-ph.GA]} \BibitemShut {NoStop}%
\bibitem [{\citenamefont {Spergel}\ and\ \citenamefont
  {Steinhardt}(2000)}]{Spergel_2000}%
  \BibitemOpen
  \bibfield  {author} {\bibinfo {author} {\bibfnamefont {D.~N.}\ \bibnamefont
  {Spergel}}\ and\ \bibinfo {author} {\bibfnamefont {P.~J.}\ \bibnamefont
  {Steinhardt}},\ }\href {\doibase 10.1103/physrevlett.84.3760} {\bibfield
  {journal} {\bibinfo  {journal} {Physical Review Letters}\ }\textbf {\bibinfo
  {volume} {84}},\ \bibinfo {pages} {3760–3763} (\bibinfo {year}
  {2000})}\BibitemShut {NoStop}%
\bibitem [{\citenamefont {Vogelsberger}\ \emph {et~al.}(2012)\citenamefont
  {Vogelsberger}, \citenamefont {Zavala},\ and\ \citenamefont
  {Loeb}}]{Vogelsberger:2012ku}%
  \BibitemOpen
  \bibfield  {author} {\bibinfo {author} {\bibfnamefont {M.}~\bibnamefont
  {Vogelsberger}}, \bibinfo {author} {\bibfnamefont {J.}~\bibnamefont
  {Zavala}}, \ and\ \bibinfo {author} {\bibfnamefont {A.}~\bibnamefont
  {Loeb}},\ }\href {\doibase 10.1111/j.1365-2966.2012.21182.x} {\bibfield
  {journal} {\bibinfo  {journal} {Mon. Not. Roy. Astron. Soc.}\ }\textbf
  {\bibinfo {volume} {423}},\ \bibinfo {pages} {3740} (\bibinfo {year}
  {2012})},\ \Eprint {http://arxiv.org/abs/1201.5892} {arXiv:1201.5892
  [astro-ph.CO]} \BibitemShut {NoStop}%
\bibitem [{\citenamefont {Rocha}\ \emph {et~al.}(2013)\citenamefont {Rocha},
  \citenamefont {Peter}, \citenamefont {Bullock}, \citenamefont {Kaplinghat},
  \citenamefont {Garrison-Kimmel}, \citenamefont {Onorbe},\ and\ \citenamefont
  {Moustakas}}]{Rocha:2012jg}%
  \BibitemOpen
  \bibfield  {author} {\bibinfo {author} {\bibfnamefont {M.}~\bibnamefont
  {Rocha}}, \bibinfo {author} {\bibfnamefont {A.~H.}\ \bibnamefont {Peter}},
  \bibinfo {author} {\bibfnamefont {J.~S.}\ \bibnamefont {Bullock}}, \bibinfo
  {author} {\bibfnamefont {M.}~\bibnamefont {Kaplinghat}}, \bibinfo {author}
  {\bibfnamefont {S.}~\bibnamefont {Garrison-Kimmel}}, \bibinfo {author}
  {\bibfnamefont {J.}~\bibnamefont {Onorbe}}, \ and\ \bibinfo {author}
  {\bibfnamefont {L.~A.}\ \bibnamefont {Moustakas}},\ }\href {\doibase
  10.1093/mnras/sts514} {\bibfield  {journal} {\bibinfo  {journal} {Mon. Not.
  Roy. Astron. Soc.}\ }\textbf {\bibinfo {volume} {430}},\ \bibinfo {pages}
  {81} (\bibinfo {year} {2013})},\ \Eprint {http://arxiv.org/abs/1208.3025}
  {arXiv:1208.3025 [astro-ph.CO]} \BibitemShut {NoStop}%
\bibitem [{\citenamefont {Peter}\ \emph {et~al.}(2013)\citenamefont {Peter},
  \citenamefont {Rocha}, \citenamefont {Bullock},\ and\ \citenamefont
  {Kaplinghat}}]{Peter:2012jh}%
  \BibitemOpen
  \bibfield  {author} {\bibinfo {author} {\bibfnamefont {A.~H.}\ \bibnamefont
  {Peter}}, \bibinfo {author} {\bibfnamefont {M.}~\bibnamefont {Rocha}},
  \bibinfo {author} {\bibfnamefont {J.~S.}\ \bibnamefont {Bullock}}, \ and\
  \bibinfo {author} {\bibfnamefont {M.}~\bibnamefont {Kaplinghat}},\ }\href
  {\doibase 10.1093/mnras/sts535} {\bibfield  {journal} {\bibinfo  {journal}
  {Mon. Not. Roy. Astron. Soc.}\ }\textbf {\bibinfo {volume} {430}},\ \bibinfo
  {pages} {105} (\bibinfo {year} {2013})},\ \Eprint
  {http://arxiv.org/abs/1208.3026} {arXiv:1208.3026 [astro-ph.CO]} \BibitemShut
  {NoStop}%
\bibitem [{\citenamefont {Zavala}\ \emph {et~al.}(2013)\citenamefont {Zavala},
  \citenamefont {Vogelsberger},\ and\ \citenamefont {Walker}}]{Zavala:2012us}%
  \BibitemOpen
  \bibfield  {author} {\bibinfo {author} {\bibfnamefont {J.}~\bibnamefont
  {Zavala}}, \bibinfo {author} {\bibfnamefont {M.}~\bibnamefont
  {Vogelsberger}}, \ and\ \bibinfo {author} {\bibfnamefont {M.~G.}\
  \bibnamefont {Walker}},\ }\href {\doibase 10.1093/mnrasl/sls053} {\bibfield
  {journal} {\bibinfo  {journal} {Mon. Not. Roy. Astron. Soc.}\ }\textbf
  {\bibinfo {volume} {431}},\ \bibinfo {pages} {L20} (\bibinfo {year}
  {2013})},\ \Eprint {http://arxiv.org/abs/1211.6426} {arXiv:1211.6426
  [astro-ph.CO]} \BibitemShut {NoStop}%
\bibitem [{\citenamefont {Elbert}\ \emph {et~al.}(2015)\citenamefont {Elbert},
  \citenamefont {Bullock}, \citenamefont {Garrison-Kimmel}, \citenamefont
  {Rocha}, \citenamefont {Oñorbe},\ and\ \citenamefont
  {Peter}}]{Elbert:2014bma}%
  \BibitemOpen
  \bibfield  {author} {\bibinfo {author} {\bibfnamefont {O.~D.}\ \bibnamefont
  {Elbert}}, \bibinfo {author} {\bibfnamefont {J.~S.}\ \bibnamefont {Bullock}},
  \bibinfo {author} {\bibfnamefont {S.}~\bibnamefont {Garrison-Kimmel}},
  \bibinfo {author} {\bibfnamefont {M.}~\bibnamefont {Rocha}}, \bibinfo
  {author} {\bibfnamefont {J.}~\bibnamefont {Oñorbe}}, \ and\ \bibinfo
  {author} {\bibfnamefont {A.~H.}\ \bibnamefont {Peter}},\ }\href {\doibase
  10.1093/mnras/stv1470} {\bibfield  {journal} {\bibinfo  {journal} {Mon. Not.
  Roy. Astron. Soc.}\ }\textbf {\bibinfo {volume} {453}},\ \bibinfo {pages}
  {29} (\bibinfo {year} {2015})},\ \Eprint {http://arxiv.org/abs/1412.1477}
  {arXiv:1412.1477 [astro-ph.GA]} \BibitemShut {NoStop}%
\bibitem [{\citenamefont {Fry}\ \emph {et~al.}(2015)\citenamefont {Fry},
  \citenamefont {Governato}, \citenamefont {Pontzen}, \citenamefont {Quinn},
  \citenamefont {Tremmel}, \citenamefont {Anderson}, \citenamefont {Menon},
  \citenamefont {Brooks},\ and\ \citenamefont {Wadsley}}]{Fry:2015rta}%
  \BibitemOpen
  \bibfield  {author} {\bibinfo {author} {\bibfnamefont {A.~B.}\ \bibnamefont
  {Fry}}, \bibinfo {author} {\bibfnamefont {F.}~\bibnamefont {Governato}},
  \bibinfo {author} {\bibfnamefont {A.}~\bibnamefont {Pontzen}}, \bibinfo
  {author} {\bibfnamefont {T.}~\bibnamefont {Quinn}}, \bibinfo {author}
  {\bibfnamefont {M.}~\bibnamefont {Tremmel}}, \bibinfo {author} {\bibfnamefont
  {L.}~\bibnamefont {Anderson}}, \bibinfo {author} {\bibfnamefont
  {H.}~\bibnamefont {Menon}}, \bibinfo {author} {\bibfnamefont
  {A.}~\bibnamefont {Brooks}}, \ and\ \bibinfo {author} {\bibfnamefont
  {J.}~\bibnamefont {Wadsley}},\ }\href {\doibase 10.1093/mnras/stv1330}
  {\bibfield  {journal} {\bibinfo  {journal} {Mon. Not. Roy. Astron. Soc.}\
  }\textbf {\bibinfo {volume} {452}},\ \bibinfo {pages} {1468} (\bibinfo {year}
  {2015})},\ \Eprint {http://arxiv.org/abs/1501.00497} {arXiv:1501.00497
  [astro-ph.CO]} \BibitemShut {NoStop}%
\bibitem [{Note1()}]{Note1}%
  \BibitemOpen
  \bibinfo {note} {For high surface brightness galaxies, the inner halo may be
  cuspy in the SIDM model due to the gravitational potential from baryons~\cite
  {Kaplinghat_2014,Vogelsberger_2014,Kaplinghat_2016,Kamada_2017,Creasey_2017,Elbert_2018,Robertson_2018,Ren_2019}.
  This behavior allows the SIDM model to solve the diversity problem~\cite
  {Oman_2015,Oman_2016}.}\BibitemShut {Stop}%
\bibitem [{\citenamefont {Markevitch}\ \emph {et~al.}(2004)\citenamefont
  {Markevitch}, \citenamefont {Gonzalez}, \citenamefont {Clowe}, \citenamefont
  {Vikhlinin}, \citenamefont {Forman}, \citenamefont {Jones}, \citenamefont
  {Murray},\ and\ \citenamefont {Tucker}}]{Markevitch_2004}%
  \BibitemOpen
  \bibfield  {author} {\bibinfo {author} {\bibfnamefont {M.}~\bibnamefont
  {Markevitch}}, \bibinfo {author} {\bibfnamefont {A.~H.}\ \bibnamefont
  {Gonzalez}}, \bibinfo {author} {\bibfnamefont {D.}~\bibnamefont {Clowe}},
  \bibinfo {author} {\bibfnamefont {A.}~\bibnamefont {Vikhlinin}}, \bibinfo
  {author} {\bibfnamefont {W.}~\bibnamefont {Forman}}, \bibinfo {author}
  {\bibfnamefont {C.}~\bibnamefont {Jones}}, \bibinfo {author} {\bibfnamefont
  {S.}~\bibnamefont {Murray}}, \ and\ \bibinfo {author} {\bibfnamefont
  {W.}~\bibnamefont {Tucker}},\ }\href {\doibase 10.1086/383178} {\bibfield
  {journal} {\bibinfo  {journal} {The Astrophysical Journal}\ }\textbf
  {\bibinfo {volume} {606}},\ \bibinfo {pages} {819–824} (\bibinfo {year}
  {2004})}\BibitemShut {NoStop}%
\bibitem [{\citenamefont {Randall}\ \emph {et~al.}(2008)\citenamefont
  {Randall}, \citenamefont {Markevitch}, \citenamefont {Clowe}, \citenamefont
  {Gonzalez},\ and\ \citenamefont {Bradač}}]{Randall_2008}%
  \BibitemOpen
  \bibfield  {author} {\bibinfo {author} {\bibfnamefont {S.~W.}\ \bibnamefont
  {Randall}}, \bibinfo {author} {\bibfnamefont {M.}~\bibnamefont {Markevitch}},
  \bibinfo {author} {\bibfnamefont {D.}~\bibnamefont {Clowe}}, \bibinfo
  {author} {\bibfnamefont {A.~H.}\ \bibnamefont {Gonzalez}}, \ and\ \bibinfo
  {author} {\bibfnamefont {M.}~\bibnamefont {Bradač}},\ }\href {\doibase
  10.1086/587859} {\bibfield  {journal} {\bibinfo  {journal} {The Astrophysical
  Journal}\ }\textbf {\bibinfo {volume} {679}},\ \bibinfo {pages} {1173–1180}
  (\bibinfo {year} {2008})}\BibitemShut {NoStop}%
\bibitem [{\citenamefont {Kahlhoefer}\ \emph {et~al.}(2013)\citenamefont
  {Kahlhoefer}, \citenamefont {Schmidt-Hoberg}, \citenamefont {Frandsen},\ and\
  \citenamefont {Sarkar}}]{Kahlhoefer_2013}%
  \BibitemOpen
  \bibfield  {author} {\bibinfo {author} {\bibfnamefont {F.}~\bibnamefont
  {Kahlhoefer}}, \bibinfo {author} {\bibfnamefont {K.}~\bibnamefont
  {Schmidt-Hoberg}}, \bibinfo {author} {\bibfnamefont {M.~T.}\ \bibnamefont
  {Frandsen}}, \ and\ \bibinfo {author} {\bibfnamefont {S.}~\bibnamefont
  {Sarkar}},\ }\href {\doibase 10.1093/mnras/stt2097} {\bibfield  {journal}
  {\bibinfo  {journal} {Monthly Notices of the Royal Astronomical Society}\
  }\textbf {\bibinfo {volume} {437}},\ \bibinfo {pages} {2865–2881} (\bibinfo
  {year} {2013})}\BibitemShut {NoStop}%
\bibitem [{\citenamefont {Harvey}\ \emph {et~al.}(2015)\citenamefont {Harvey},
  \citenamefont {Massey}, \citenamefont {Kitching}, \citenamefont {Taylor},\
  and\ \citenamefont {Tittley}}]{Harvey_2015}%
  \BibitemOpen
  \bibfield  {author} {\bibinfo {author} {\bibfnamefont {D.}~\bibnamefont
  {Harvey}}, \bibinfo {author} {\bibfnamefont {R.}~\bibnamefont {Massey}},
  \bibinfo {author} {\bibfnamefont {T.}~\bibnamefont {Kitching}}, \bibinfo
  {author} {\bibfnamefont {A.}~\bibnamefont {Taylor}}, \ and\ \bibinfo {author}
  {\bibfnamefont {E.}~\bibnamefont {Tittley}},\ }\href {\doibase
  10.1126/science.1261381} {\bibfield  {journal} {\bibinfo  {journal}
  {Science}\ }\textbf {\bibinfo {volume} {347}},\ \bibinfo {pages}
  {1462–1465} (\bibinfo {year} {2015})}\BibitemShut {NoStop}%
\bibitem [{\citenamefont {Robertson}\ \emph {et~al.}(2016)\citenamefont
  {Robertson}, \citenamefont {Massey},\ and\ \citenamefont
  {Eke}}]{Robertson_2016}%
  \BibitemOpen
  \bibfield  {author} {\bibinfo {author} {\bibfnamefont {A.}~\bibnamefont
  {Robertson}}, \bibinfo {author} {\bibfnamefont {R.}~\bibnamefont {Massey}}, \
  and\ \bibinfo {author} {\bibfnamefont {V.}~\bibnamefont {Eke}},\ }\href
  {\doibase 10.1093/mnras/stw2670} {\bibfield  {journal} {\bibinfo  {journal}
  {Monthly Notices of the Royal Astronomical Society}\ }\textbf {\bibinfo
  {volume} {465}},\ \bibinfo {pages} {569–587} (\bibinfo {year}
  {2016})}\BibitemShut {NoStop}%
\bibitem [{\citenamefont {Wittman}\ \emph {et~al.}(2018)\citenamefont
  {Wittman}, \citenamefont {Golovich},\ and\ \citenamefont
  {Dawson}}]{Wittman_2018}%
  \BibitemOpen
  \bibfield  {author} {\bibinfo {author} {\bibfnamefont {D.}~\bibnamefont
  {Wittman}}, \bibinfo {author} {\bibfnamefont {N.}~\bibnamefont {Golovich}}, \
  and\ \bibinfo {author} {\bibfnamefont {W.~A.}\ \bibnamefont {Dawson}},\
  }\href {\doibase 10.3847/1538-4357/aaee77} {\bibfield  {journal} {\bibinfo
  {journal} {The Astrophysical Journal}\ }\textbf {\bibinfo {volume} {869}},\
  \bibinfo {pages} {104} (\bibinfo {year} {2018})}\BibitemShut {NoStop}%
\bibitem [{\citenamefont {Harvey}\ \emph {et~al.}(2019)\citenamefont {Harvey},
  \citenamefont {Robertson}, \citenamefont {Massey},\ and\ \citenamefont
  {McCarthy}}]{Harvey_2019}%
  \BibitemOpen
  \bibfield  {author} {\bibinfo {author} {\bibfnamefont {D.}~\bibnamefont
  {Harvey}}, \bibinfo {author} {\bibfnamefont {A.}~\bibnamefont {Robertson}},
  \bibinfo {author} {\bibfnamefont {R.}~\bibnamefont {Massey}}, \ and\ \bibinfo
  {author} {\bibfnamefont {I.~G.}\ \bibnamefont {McCarthy}},\ }\href {\doibase
  10.1093/mnras/stz1816} {\bibfield  {journal} {\bibinfo  {journal} {Monthly
  Notices of the Royal Astronomical Society}\ }\textbf {\bibinfo {volume}
  {488}},\ \bibinfo {pages} {1572–1579} (\bibinfo {year} {2019})}\BibitemShut
  {NoStop}%
\bibitem [{\citenamefont {Bondarenko}\ \emph {et~al.}(2020)\citenamefont
  {Bondarenko}, \citenamefont {Sokolenko}, \citenamefont {Boyarsky},
  \citenamefont {Robertson}, \citenamefont {Harvey},\ and\ \citenamefont
  {Revaz}}]{Bondarenko:2020mpf}%
  \BibitemOpen
  \bibfield  {author} {\bibinfo {author} {\bibfnamefont {K.}~\bibnamefont
  {Bondarenko}}, \bibinfo {author} {\bibfnamefont {A.}~\bibnamefont
  {Sokolenko}}, \bibinfo {author} {\bibfnamefont {A.}~\bibnamefont {Boyarsky}},
  \bibinfo {author} {\bibfnamefont {A.}~\bibnamefont {Robertson}}, \bibinfo
  {author} {\bibfnamefont {D.}~\bibnamefont {Harvey}}, \ and\ \bibinfo {author}
  {\bibfnamefont {Y.}~\bibnamefont {Revaz}},\ }\href@noop {} {\  (\bibinfo
  {year} {2020})},\ \Eprint {http://arxiv.org/abs/2006.06623} {arXiv:2006.06623
  [astro-ph.CO]} \BibitemShut {NoStop}%
\bibitem [{\citenamefont {Sagunski}\ \emph {et~al.}(2020)\citenamefont
  {Sagunski}, \citenamefont {Gad-Nasr}, \citenamefont {Colquhoun},
  \citenamefont {Robertson},\ and\ \citenamefont {Tulin}}]{Sagunski:2020spe}%
  \BibitemOpen
  \bibfield  {author} {\bibinfo {author} {\bibfnamefont {L.}~\bibnamefont
  {Sagunski}}, \bibinfo {author} {\bibfnamefont {S.}~\bibnamefont {Gad-Nasr}},
  \bibinfo {author} {\bibfnamefont {B.}~\bibnamefont {Colquhoun}}, \bibinfo
  {author} {\bibfnamefont {A.}~\bibnamefont {Robertson}}, \ and\ \bibinfo
  {author} {\bibfnamefont {S.}~\bibnamefont {Tulin}},\ }\href@noop {} {\
  (\bibinfo {year} {2020})},\ \Eprint {http://arxiv.org/abs/2006.12515}
  {arXiv:2006.12515 [astro-ph.CO]} \BibitemShut {NoStop}%
\bibitem [{\citenamefont {Read}\ and\ \citenamefont
  {Gilmore}(2005)}]{Read:2004xc}%
  \BibitemOpen
  \bibfield  {author} {\bibinfo {author} {\bibfnamefont {J.~I.}\ \bibnamefont
  {Read}}\ and\ \bibinfo {author} {\bibfnamefont {G.}~\bibnamefont {Gilmore}},\
  }\href {\doibase 10.1111/j.1365-2966.2004.08424.x} {\bibfield  {journal}
  {\bibinfo  {journal} {Mon. Not. Roy. Astron. Soc.}\ }\textbf {\bibinfo
  {volume} {356}},\ \bibinfo {pages} {107} (\bibinfo {year} {2005})},\ \Eprint
  {http://arxiv.org/abs/astro-ph/0409565} {arXiv:astro-ph/0409565} \BibitemShut
  {NoStop}%
\bibitem [{\citenamefont {Mashchenko}\ \emph {et~al.}(2008)\citenamefont
  {Mashchenko}, \citenamefont {Wadsley},\ and\ \citenamefont
  {Couchman}}]{Mashchenko:2007jp}%
  \BibitemOpen
  \bibfield  {author} {\bibinfo {author} {\bibfnamefont {S.}~\bibnamefont
  {Mashchenko}}, \bibinfo {author} {\bibfnamefont {J.}~\bibnamefont {Wadsley}},
  \ and\ \bibinfo {author} {\bibfnamefont {H.}~\bibnamefont {Couchman}},\
  }\href {\doibase 10.1126/science.1148666} {\bibfield  {journal} {\bibinfo
  {journal} {Science}\ }\textbf {\bibinfo {volume} {319}},\ \bibinfo {pages}
  {174} (\bibinfo {year} {2008})},\ \Eprint {http://arxiv.org/abs/0711.4803}
  {arXiv:0711.4803 [astro-ph]} \BibitemShut {NoStop}%
\bibitem [{\citenamefont {Pontzen}\ and\ \citenamefont
  {Governato}(2012)}]{Pontzen:2011ty}%
  \BibitemOpen
  \bibfield  {author} {\bibinfo {author} {\bibfnamefont {A.}~\bibnamefont
  {Pontzen}}\ and\ \bibinfo {author} {\bibfnamefont {F.}~\bibnamefont
  {Governato}},\ }\href {\doibase 10.1111/j.1365-2966.2012.20571.x} {\bibfield
  {journal} {\bibinfo  {journal} {Mon. Not. Roy. Astron. Soc.}\ }\textbf
  {\bibinfo {volume} {421}},\ \bibinfo {pages} {3464} (\bibinfo {year}
  {2012})},\ \Eprint {http://arxiv.org/abs/1106.0499} {arXiv:1106.0499
  [astro-ph.CO]} \BibitemShut {NoStop}%
\bibitem [{\citenamefont {Oñorbe}\ \emph {et~al.}(2015)\citenamefont
  {Oñorbe}, \citenamefont {Boylan-Kolchin}, \citenamefont {Bullock},
  \citenamefont {Hopkins}, \citenamefont {Ker\v~es}, \citenamefont
  {Faucher-Giguère}, \citenamefont {Quataert},\ and\ \citenamefont
  {Murray}}]{Onorbe:2015ija}%
  \BibitemOpen
  \bibfield  {author} {\bibinfo {author} {\bibfnamefont {J.}~\bibnamefont
  {Oñorbe}}, \bibinfo {author} {\bibfnamefont {M.}~\bibnamefont
  {Boylan-Kolchin}}, \bibinfo {author} {\bibfnamefont {J.~S.}\ \bibnamefont
  {Bullock}}, \bibinfo {author} {\bibfnamefont {P.~F.}\ \bibnamefont
  {Hopkins}}, \bibinfo {author} {\bibfnamefont {D.~s.}\ \bibnamefont
  {Ker\v~es}}, \bibinfo {author} {\bibfnamefont {C.-A.}\ \bibnamefont
  {Faucher-Giguère}}, \bibinfo {author} {\bibfnamefont {E.}~\bibnamefont
  {Quataert}}, \ and\ \bibinfo {author} {\bibfnamefont {N.}~\bibnamefont
  {Murray}},\ }\href {\doibase 10.1093/mnras/stv2072} {\bibfield  {journal}
  {\bibinfo  {journal} {Mon. Not. Roy. Astron. Soc.}\ }\textbf {\bibinfo
  {volume} {454}},\ \bibinfo {pages} {2092} (\bibinfo {year} {2015})},\ \Eprint
  {http://arxiv.org/abs/1502.02036} {arXiv:1502.02036 [astro-ph.GA]}
  \BibitemShut {NoStop}%
\bibitem [{\citenamefont {Tollet}\ \emph {et~al.}(2016)\citenamefont {Tollet}
  \emph {et~al.}}]{Tollet:2015gqa}%
  \BibitemOpen
  \bibfield  {author} {\bibinfo {author} {\bibfnamefont {E.}~\bibnamefont
  {Tollet}} \emph {et~al.},\ }\href {\doibase 10.1093/mnras/stv2856} {\bibfield
   {journal} {\bibinfo  {journal} {Mon. Not. Roy. Astron. Soc.}\ }\textbf
  {\bibinfo {volume} {456}},\ \bibinfo {pages} {3542} (\bibinfo {year}
  {2016})},\ \bibinfo {note} {[Erratum: Mon.Not.Roy.Astron.Soc. 487, 1764
  (2019)]},\ \Eprint {http://arxiv.org/abs/1507.03590} {arXiv:1507.03590
  [astro-ph.GA]} \BibitemShut {NoStop}%
\bibitem [{\citenamefont {Hayashi}\ \emph {et~al.}(2020)\citenamefont
  {Hayashi}, \citenamefont {Chiba},\ and\ \citenamefont
  {Ishiyama}}]{Hayashi:2020jze}%
  \BibitemOpen
  \bibfield  {author} {\bibinfo {author} {\bibfnamefont {K.}~\bibnamefont
  {Hayashi}}, \bibinfo {author} {\bibfnamefont {M.}~\bibnamefont {Chiba}}, \
  and\ \bibinfo {author} {\bibfnamefont {T.}~\bibnamefont {Ishiyama}},\
  }\href@noop {} {\  (\bibinfo {year} {2020})},\ \Eprint
  {http://arxiv.org/abs/2007.13780} {arXiv:2007.13780 [astro-ph.GA]}
  \BibitemShut {NoStop}%
\bibitem [{\citenamefont {Kaplinghat}\ \emph {et~al.}(2016)\citenamefont
  {Kaplinghat}, \citenamefont {Tulin},\ and\ \citenamefont
  {Yu}}]{Kaplinghat_2016}%
  \BibitemOpen
  \bibfield  {author} {\bibinfo {author} {\bibfnamefont {M.}~\bibnamefont
  {Kaplinghat}}, \bibinfo {author} {\bibfnamefont {S.}~\bibnamefont {Tulin}}, \
  and\ \bibinfo {author} {\bibfnamefont {H.-B.}\ \bibnamefont {Yu}},\ }\href
  {\doibase 10.1103/physrevlett.116.041302} {\bibfield  {journal} {\bibinfo
  {journal} {Physical Review Letters}\ }\textbf {\bibinfo {volume} {116}}
  (\bibinfo {year} {2016}),\ 10.1103/physrevlett.116.041302}\BibitemShut
  {NoStop}%
\bibitem [{\citenamefont {Valli}\ and\ \citenamefont
  {Yu}(2018)}]{Valli:2017ktb}%
  \BibitemOpen
  \bibfield  {author} {\bibinfo {author} {\bibfnamefont {M.}~\bibnamefont
  {Valli}}\ and\ \bibinfo {author} {\bibfnamefont {H.-B.}\ \bibnamefont {Yu}},\
  }\href {\doibase 10.1038/s41550-018-0560-7} {\bibfield  {journal} {\bibinfo
  {journal} {Nature Astron.}\ }\textbf {\bibinfo {volume} {2}},\ \bibinfo
  {pages} {907} (\bibinfo {year} {2018})},\ \Eprint
  {http://arxiv.org/abs/1711.03502} {arXiv:1711.03502 [astro-ph.GA]}
  \BibitemShut {NoStop}%
\bibitem [{\citenamefont {{Binney}}\ and\ \citenamefont
  {{Tremaine}}(2008)}]{2008gady.book.....B}%
  \BibitemOpen
  \bibfield  {author} {\bibinfo {author} {\bibfnamefont {J.}~\bibnamefont
  {{Binney}}}\ and\ \bibinfo {author} {\bibfnamefont {S.}~\bibnamefont
  {{Tremaine}}},\ }\href@noop {} {\emph {\bibinfo {title} {{Galactic Dynamics:
  Second Edition}}}}\ (\bibinfo {year} {2008})\BibitemShut {NoStop}%
\bibitem [{\citenamefont {Baes}\ and\ \citenamefont
  {Van~Hese}(2007)}]{Baes:2007tx}%
  \BibitemOpen
  \bibfield  {author} {\bibinfo {author} {\bibfnamefont {M.}~\bibnamefont
  {Baes}}\ and\ \bibinfo {author} {\bibfnamefont {E.}~\bibnamefont
  {Van~Hese}},\ }\href {\doibase 10.1051/0004-6361:20077672} {\bibfield
  {journal} {\bibinfo  {journal} {Astron. Astrophys.}\ }\textbf {\bibinfo
  {volume} {471}},\ \bibinfo {pages} {419} (\bibinfo {year} {2007})},\ \Eprint
  {http://arxiv.org/abs/0705.4109} {arXiv:0705.4109 [astro-ph]} \BibitemShut
  {NoStop}%
\bibitem [{\citenamefont {Plummer}(1911)}]{Plummer:1911zza}%
  \BibitemOpen
  \bibfield  {author} {\bibinfo {author} {\bibfnamefont {H.~C.}\ \bibnamefont
  {Plummer}},\ }\href@noop {} {\bibfield  {journal} {\bibinfo  {journal} {Mon.
  Not. Roy. Astron. Soc.}\ }\textbf {\bibinfo {volume} {71}},\ \bibinfo {pages}
  {460} (\bibinfo {year} {1911})}\BibitemShut {NoStop}%
\bibitem [{Note2()}]{Note2}%
  \BibitemOpen
  \bibinfo {note} {Here, the mass density and the pressure of DM are related by
  $p_{DM}=\sigma _0^2\rho _{DM}$. This corresponds to the Maxwell-Boltzmann
  velocity distribution, $f_{DM} \propto e^{-v^2/2\sigma _0^2}$. We do not
  truncate the distribution by the escape velocity. In this case, the mean
  relative DM velocity is given by, $v = 4\sigma _0/\protect \sqrt {\pi
  }$.}\BibitemShut {Stop}%
\bibitem [{\citenamefont {Navarro}\ \emph {et~al.}(1997)\citenamefont
  {Navarro}, \citenamefont {Frenk},\ and\ \citenamefont
  {White}}]{Navarro:1996gj}%
  \BibitemOpen
  \bibfield  {author} {\bibinfo {author} {\bibfnamefont {J.~F.}\ \bibnamefont
  {Navarro}}, \bibinfo {author} {\bibfnamefont {C.~S.}\ \bibnamefont {Frenk}},
  \ and\ \bibinfo {author} {\bibfnamefont {S.~D.}\ \bibnamefont {White}},\
  }\href {\doibase 10.1086/304888} {\bibfield  {journal} {\bibinfo  {journal}
  {Astrophys. J.}\ }\textbf {\bibinfo {volume} {490}},\ \bibinfo {pages} {493}
  (\bibinfo {year} {1997})},\ \Eprint {http://arxiv.org/abs/astro-ph/9611107}
  {arXiv:astro-ph/9611107} \BibitemShut {NoStop}%
\bibitem [{\citenamefont {Vogelsberger}\ \emph {et~al.}(2016)\citenamefont
  {Vogelsberger}, \citenamefont {Zavala}, \citenamefont {Cyr-Racine},
  \citenamefont {Pfrommer}, \citenamefont {Bringmann},\ and\ \citenamefont
  {Sigurdson}}]{Vogelsberger:2015gpr}%
  \BibitemOpen
  \bibfield  {author} {\bibinfo {author} {\bibfnamefont {M.}~\bibnamefont
  {Vogelsberger}}, \bibinfo {author} {\bibfnamefont {J.}~\bibnamefont
  {Zavala}}, \bibinfo {author} {\bibfnamefont {F.-Y.}\ \bibnamefont
  {Cyr-Racine}}, \bibinfo {author} {\bibfnamefont {C.}~\bibnamefont
  {Pfrommer}}, \bibinfo {author} {\bibfnamefont {T.}~\bibnamefont {Bringmann}},
  \ and\ \bibinfo {author} {\bibfnamefont {K.}~\bibnamefont {Sigurdson}},\
  }\href {\doibase 10.1093/mnras/stw1076} {\bibfield  {journal} {\bibinfo
  {journal} {Mon. Not. Roy. Astron. Soc.}\ }\textbf {\bibinfo {volume} {460}},\
  \bibinfo {pages} {1399} (\bibinfo {year} {2016})},\ \Eprint
  {http://arxiv.org/abs/1512.05349} {arXiv:1512.05349 [astro-ph.CO]}
  \BibitemShut {NoStop}%
\bibitem [{\citenamefont {Boylan-Kolchin}\ \emph {et~al.}(2011)\citenamefont
  {Boylan-Kolchin}, \citenamefont {Bullock},\ and\ \citenamefont
  {Kaplinghat}}]{BoylanKolchin:2011de}%
  \BibitemOpen
  \bibfield  {author} {\bibinfo {author} {\bibfnamefont {M.}~\bibnamefont
  {Boylan-Kolchin}}, \bibinfo {author} {\bibfnamefont {J.~S.}\ \bibnamefont
  {Bullock}}, \ and\ \bibinfo {author} {\bibfnamefont {M.}~\bibnamefont
  {Kaplinghat}},\ }\href {\doibase 10.1111/j.1745-3933.2011.01074.x} {\bibfield
   {journal} {\bibinfo  {journal} {Mon. Not. Roy. Astron. Soc.}\ }\textbf
  {\bibinfo {volume} {415}},\ \bibinfo {pages} {L40} (\bibinfo {year}
  {2011})},\ \Eprint {http://arxiv.org/abs/1103.0007} {arXiv:1103.0007
  [astro-ph.CO]} \BibitemShut {NoStop}%
\bibitem [{\citenamefont {Boylan-Kolchin}\ \emph {et~al.}(2012)\citenamefont
  {Boylan-Kolchin}, \citenamefont {Bullock},\ and\ \citenamefont
  {Kaplinghat}}]{BoylanKolchin:2011dk}%
  \BibitemOpen
  \bibfield  {author} {\bibinfo {author} {\bibfnamefont {M.}~\bibnamefont
  {Boylan-Kolchin}}, \bibinfo {author} {\bibfnamefont {J.~S.}\ \bibnamefont
  {Bullock}}, \ and\ \bibinfo {author} {\bibfnamefont {M.}~\bibnamefont
  {Kaplinghat}},\ }\href {\doibase 10.1111/j.1365-2966.2012.20695.x} {\bibfield
   {journal} {\bibinfo  {journal} {Mon. Not. Roy. Astron. Soc.}\ }\textbf
  {\bibinfo {volume} {422}},\ \bibinfo {pages} {1203} (\bibinfo {year}
  {2012})},\ \Eprint {http://arxiv.org/abs/1111.2048} {arXiv:1111.2048
  [astro-ph.CO]} \BibitemShut {NoStop}%
\bibitem [{\citenamefont {Molin\'e}\ \emph {et~al.}(2017)\citenamefont
  {Molin\'e}, \citenamefont {S\'anchez-Conde}, \citenamefont {Palomares-Ruiz},\
  and\ \citenamefont {Prada}}]{Moline:2016pbm}%
  \BibitemOpen
  \bibfield  {author} {\bibinfo {author} {\bibfnamefont {A.}~\bibnamefont
  {Molin\'e}}, \bibinfo {author} {\bibfnamefont {M.~A.}\ \bibnamefont
  {S\'anchez-Conde}}, \bibinfo {author} {\bibfnamefont {S.}~\bibnamefont
  {Palomares-Ruiz}}, \ and\ \bibinfo {author} {\bibfnamefont {F.}~\bibnamefont
  {Prada}},\ }\href {\doibase 10.1093/mnras/stx026} {\bibfield  {journal}
  {\bibinfo  {journal} {Mon. Not. Roy. Astron. Soc.}\ }\textbf {\bibinfo
  {volume} {466}},\ \bibinfo {pages} {4974} (\bibinfo {year} {2017})},\ \Eprint
  {http://arxiv.org/abs/1603.04057} {arXiv:1603.04057 [astro-ph.CO]}
  \BibitemShut {NoStop}%
\bibitem [{\citenamefont {Ishiyama}\ and\ \citenamefont
  {Ando}(2020)}]{Ishiyama:2019hmh}%
  \BibitemOpen
  \bibfield  {author} {\bibinfo {author} {\bibfnamefont {T.}~\bibnamefont
  {Ishiyama}}\ and\ \bibinfo {author} {\bibfnamefont {S.}~\bibnamefont
  {Ando}},\ }\href {\doibase 10.1093/mnras/staa069} {\bibfield  {journal}
  {\bibinfo  {journal} {Mon. Not. Roy. Astron. Soc.}\ }\textbf {\bibinfo
  {volume} {492}},\ \bibinfo {pages} {3662} (\bibinfo {year} {2020})},\ \Eprint
  {http://arxiv.org/abs/1907.03642} {arXiv:1907.03642 [astro-ph.CO]}
  \BibitemShut {NoStop}%
\bibitem [{\citenamefont {Wang}\ \emph {et~al.}(2020)\citenamefont {Wang},
  \citenamefont {Han}, \citenamefont {Cautun}, \citenamefont {Li},\ and\
  \citenamefont {Ishigaki}}]{Wang:2019ubx}%
  \BibitemOpen
  \bibfield  {author} {\bibinfo {author} {\bibfnamefont {W.}~\bibnamefont
  {Wang}}, \bibinfo {author} {\bibfnamefont {J.}~\bibnamefont {Han}}, \bibinfo
  {author} {\bibfnamefont {M.}~\bibnamefont {Cautun}}, \bibinfo {author}
  {\bibfnamefont {Z.}~\bibnamefont {Li}}, \ and\ \bibinfo {author}
  {\bibfnamefont {M.~N.}\ \bibnamefont {Ishigaki}},\ }\href {\doibase
  10.1007/s11433-019-1541-6} {\bibfield  {journal} {\bibinfo  {journal} {Sci.
  China Phys. Mech. Astron.}\ }\textbf {\bibinfo {volume} {63}},\ \bibinfo
  {pages} {109801} (\bibinfo {year} {2020})},\ \Eprint
  {http://arxiv.org/abs/1912.02599} {arXiv:1912.02599 [astro-ph.GA]}
  \BibitemShut {NoStop}%
\bibitem [{\citenamefont {{Martin}}\ \emph {et~al.}(2008)\citenamefont
  {{Martin}}, \citenamefont {{de Jong}},\ and\ \citenamefont
  {{Rix}}}]{2008ApJ...684.1075M}%
  \BibitemOpen
  \bibfield  {author} {\bibinfo {author} {\bibfnamefont {N.~F.}\ \bibnamefont
  {{Martin}}}, \bibinfo {author} {\bibfnamefont {J.~T.~A.}\ \bibnamefont {{de
  Jong}}}, \ and\ \bibinfo {author} {\bibfnamefont {H.-W.}\ \bibnamefont
  {{Rix}}},\ }\href {\doibase 10.1086/590336} {\bibfield  {journal} {\bibinfo
  {journal} {Astrophys. J.}\ }\textbf {\bibinfo {volume} {684}},\ \bibinfo
  {pages} {1075} (\bibinfo {year} {2008})},\ \Eprint
  {http://arxiv.org/abs/0805.2945} {arXiv:0805.2945 [astro-ph]} \BibitemShut
  {NoStop}%
\bibitem [{\citenamefont {Belokurov}\ \emph {et~al.}(2009)\citenamefont
  {Belokurov}, \citenamefont {Walker}, \citenamefont {Evans}, \citenamefont
  {Gilmore}, \citenamefont {Irwin}, \citenamefont {Mateo}, \citenamefont
  {Mayer}, \citenamefont {Olszewski}, \citenamefont {Bechtold},\ and\
  \citenamefont {Pickering}}]{2009MNRAS.397.1748B}%
  \BibitemOpen
  \bibfield  {author} {\bibinfo {author} {\bibfnamefont {V.}~\bibnamefont
  {Belokurov}}, \bibinfo {author} {\bibfnamefont {M.}~\bibnamefont {Walker}},
  \bibinfo {author} {\bibfnamefont {N.}~\bibnamefont {Evans}}, \bibinfo
  {author} {\bibfnamefont {G.}~\bibnamefont {Gilmore}}, \bibinfo {author}
  {\bibfnamefont {M.}~\bibnamefont {Irwin}}, \bibinfo {author} {\bibfnamefont
  {M.}~\bibnamefont {Mateo}}, \bibinfo {author} {\bibfnamefont
  {L.}~\bibnamefont {Mayer}}, \bibinfo {author} {\bibfnamefont
  {E.}~\bibnamefont {Olszewski}}, \bibinfo {author} {\bibfnamefont
  {J.}~\bibnamefont {Bechtold}}, \ and\ \bibinfo {author} {\bibfnamefont
  {T.}~\bibnamefont {Pickering}},\ }\href {\doibase
  10.1111/j.1365-2966.2009.15106.x} {\bibfield  {journal} {\bibinfo  {journal}
  {Mon. Not. Roy. Astron. Soc.}\ }\textbf {\bibinfo {volume} {397}},\ \bibinfo
  {pages} {1748} (\bibinfo {year} {2009})},\ \Eprint
  {http://arxiv.org/abs/0903.0818} {arXiv:0903.0818 [astro-ph.GA]} \BibitemShut
  {NoStop}%
\bibitem [{\citenamefont {{Koposov}}\ \emph {et~al.}(2015)\citenamefont
  {{Koposov}}, \citenamefont {{Belokurov}}, \citenamefont {{Torrealba}},\ and\
  \citenamefont {{Evans}}}]{2015ApJ...805..130K}%
  \BibitemOpen
  \bibfield  {author} {\bibinfo {author} {\bibfnamefont {S.~E.}\ \bibnamefont
  {{Koposov}}}, \bibinfo {author} {\bibfnamefont {V.}~\bibnamefont
  {{Belokurov}}}, \bibinfo {author} {\bibfnamefont {G.}~\bibnamefont
  {{Torrealba}}}, \ and\ \bibinfo {author} {\bibfnamefont {N.~W.}\ \bibnamefont
  {{Evans}}},\ }\href {\doibase 10.1088/0004-637X/805/2/130} {\bibfield
  {journal} {\bibinfo  {journal} {Astrophys. J.}\ }\textbf {\bibinfo {volume}
  {805}},\ \bibinfo {eid} {130} (\bibinfo {year} {2015})},\ \Eprint
  {http://arxiv.org/abs/1503.02079} {arXiv:1503.02079 [astro-ph.GA]}
  \BibitemShut {NoStop}%
\bibitem [{\citenamefont {Martin}\ \emph {et~al.}(2015)\citenamefont {Martin}
  \emph {et~al.}}]{2015ApJ...804L...5M}%
  \BibitemOpen
  \bibfield  {author} {\bibinfo {author} {\bibfnamefont {N.~F.}\ \bibnamefont
  {Martin}} \emph {et~al.},\ }\href {\doibase 10.1088/2041-8205/804/1/L5}
  {\bibfield  {journal} {\bibinfo  {journal} {Astrophys. J. Lett.}\ }\textbf
  {\bibinfo {volume} {804}},\ \bibinfo {eid} {L5} (\bibinfo {year} {2015})},\
  \Eprint {http://arxiv.org/abs/1503.06216} {arXiv:1503.06216 [astro-ph.GA]}
  \BibitemShut {NoStop}%
\bibitem [{\citenamefont {Sand}\ \emph {et~al.}(2012)\citenamefont {Sand},
  \citenamefont {Strader}, \citenamefont {Willman}, \citenamefont {Zaritsky},
  \citenamefont {McLeod}, \citenamefont {Caldwell}, \citenamefont {Seth},\ and\
  \citenamefont {Olszewski}}]{2012ApJ...756...79S}%
  \BibitemOpen
  \bibfield  {author} {\bibinfo {author} {\bibfnamefont {D.}~\bibnamefont
  {Sand}}, \bibinfo {author} {\bibfnamefont {J.}~\bibnamefont {Strader}},
  \bibinfo {author} {\bibfnamefont {B.}~\bibnamefont {Willman}}, \bibinfo
  {author} {\bibfnamefont {D.}~\bibnamefont {Zaritsky}}, \bibinfo {author}
  {\bibfnamefont {B.}~\bibnamefont {McLeod}}, \bibinfo {author} {\bibfnamefont
  {N.}~\bibnamefont {Caldwell}}, \bibinfo {author} {\bibfnamefont
  {A.}~\bibnamefont {Seth}}, \ and\ \bibinfo {author} {\bibfnamefont
  {E.}~\bibnamefont {Olszewski}},\ }\href {\doibase 10.1088/0004-637X/756/1/79}
  {\bibfield  {journal} {\bibinfo  {journal} {Astrophys. J.}\ }\textbf
  {\bibinfo {volume} {756}},\ \bibinfo {eid} {79} (\bibinfo {year} {2012})},\
  \Eprint {http://arxiv.org/abs/1111.6608} {arXiv:1111.6608 [astro-ph.CO]}
  \BibitemShut {NoStop}%
\bibitem [{\citenamefont {Drlica-Wagner}\ \emph {et~al.}(2015)\citenamefont
  {Drlica-Wagner} \emph {et~al.}}]{2015ApJ...813..109D}%
  \BibitemOpen
  \bibfield  {author} {\bibinfo {author} {\bibfnamefont {A.}~\bibnamefont
  {Drlica-Wagner}} \emph {et~al.} (\bibinfo {collaboration} {DES}),\ }\href
  {\doibase 10.1088/0004-637X/813/2/109} {\bibfield  {journal} {\bibinfo
  {journal} {Astrophys. J.}\ }\textbf {\bibinfo {volume} {813}},\ \bibinfo
  {eid} {109} (\bibinfo {year} {2015})},\ \Eprint
  {http://arxiv.org/abs/1508.03622} {arXiv:1508.03622 [astro-ph.GA]}
  \BibitemShut {NoStop}%
\bibitem [{\citenamefont {Bechtol}\ \emph {et~al.}(2015)\citenamefont {Bechtol}
  \emph {et~al.}}]{2015ApJ...807...50B}%
  \BibitemOpen
  \bibfield  {author} {\bibinfo {author} {\bibfnamefont {K.}~\bibnamefont
  {Bechtol}} \emph {et~al.} (\bibinfo {collaboration} {DES}),\ }\href {\doibase
  10.1088/0004-637X/807/1/50} {\bibfield  {journal} {\bibinfo  {journal}
  {Astrophys. J.}\ }\textbf {\bibinfo {volume} {807}},\ \bibinfo {eid} {50}
  (\bibinfo {year} {2015})},\ \Eprint {http://arxiv.org/abs/1503.02584}
  {arXiv:1503.02584 [astro-ph.GA]} \BibitemShut {NoStop}%
\bibitem [{\citenamefont {{Mu{\~n}oz}}\ \emph {et~al.}(2018)\citenamefont
  {{Mu{\~n}oz}} \emph {et~al.}}]{2018ApJ...860...66M}%
  \BibitemOpen
  \bibfield  {author} {\bibinfo {author} {\bibfnamefont {R.~R.}\ \bibnamefont
  {{Mu{\~n}oz}}} \emph {et~al.},\ }\href {\doibase 10.3847/1538-4357/aac16b}
  {\bibfield  {journal} {\bibinfo  {journal} {\apj}\ }\textbf {\bibinfo
  {volume} {860}},\ \bibinfo {eid} {66} (\bibinfo {year} {2018})},\ \Eprint
  {http://arxiv.org/abs/1806.06891} {arXiv:1806.06891 [astro-ph.GA]}
  \BibitemShut {NoStop}%
\bibitem [{\citenamefont {Simon}\ \emph {et~al.}(2011)\citenamefont {Simon}
  \emph {et~al.}}]{2011ApJ...733...46S}%
  \BibitemOpen
  \bibfield  {author} {\bibinfo {author} {\bibfnamefont {J.~D.}\ \bibnamefont
  {Simon}} \emph {et~al.},\ }\href {\doibase 10.1088/0004-637X/733/1/46}
  {\bibfield  {journal} {\bibinfo  {journal} {Astrophys. J.}\ }\textbf
  {\bibinfo {volume} {733}},\ \bibinfo {eid} {46} (\bibinfo {year} {2011})},\
  \Eprint {http://arxiv.org/abs/1007.4198} {arXiv:1007.4198 [astro-ph.GA]}
  \BibitemShut {NoStop}%
\bibitem [{\citenamefont {Kirby}\ \emph {et~al.}(2013)\citenamefont {Kirby},
  \citenamefont {Boylan-Kolchin}, \citenamefont {Cohen}, \citenamefont {Geha},
  \citenamefont {Bullock},\ and\ \citenamefont
  {Kaplinghat}}]{2013ApJ...770...16K}%
  \BibitemOpen
  \bibfield  {author} {\bibinfo {author} {\bibfnamefont {E.~N.}\ \bibnamefont
  {Kirby}}, \bibinfo {author} {\bibfnamefont {M.}~\bibnamefont
  {Boylan-Kolchin}}, \bibinfo {author} {\bibfnamefont {J.~G.}\ \bibnamefont
  {Cohen}}, \bibinfo {author} {\bibfnamefont {M.}~\bibnamefont {Geha}},
  \bibinfo {author} {\bibfnamefont {J.~S.}\ \bibnamefont {Bullock}}, \ and\
  \bibinfo {author} {\bibfnamefont {M.}~\bibnamefont {Kaplinghat}},\ }\href
  {\doibase 10.1088/0004-637X/770/1/16} {\bibfield  {journal} {\bibinfo
  {journal} {Astrophys. J.}\ }\textbf {\bibinfo {volume} {770}},\ \bibinfo
  {eid} {16} (\bibinfo {year} {2013})},\ \Eprint
  {http://arxiv.org/abs/1304.6080} {arXiv:1304.6080 [astro-ph.CO]} \BibitemShut
  {NoStop}%
\bibitem [{\citenamefont {Koposov}\ \emph {et~al.}(2011)\citenamefont {Koposov}
  \emph {et~al.}}]{2011ApJ...736..146K}%
  \BibitemOpen
  \bibfield  {author} {\bibinfo {author} {\bibfnamefont {S.~E.}\ \bibnamefont
  {Koposov}} \emph {et~al.},\ }\href {\doibase 10.1088/0004-637X/736/2/146}
  {\bibfield  {journal} {\bibinfo  {journal} {Astrophys. J.}\ }\textbf
  {\bibinfo {volume} {736}},\ \bibinfo {eid} {146} (\bibinfo {year} {2011})},\
  \Eprint {http://arxiv.org/abs/1105.4102} {arXiv:1105.4102 [astro-ph.GA]}
  \BibitemShut {NoStop}%
\bibitem [{\citenamefont {{Simon}}\ and\ \citenamefont
  {{Geha}}(2007)}]{2007ApJ...670..313S}%
  \BibitemOpen
  \bibfield  {author} {\bibinfo {author} {\bibfnamefont {J.~D.}\ \bibnamefont
  {{Simon}}}\ and\ \bibinfo {author} {\bibfnamefont {M.}~\bibnamefont
  {{Geha}}},\ }\href {\doibase 10.1086/521816} {\bibfield  {journal} {\bibinfo
  {journal} {Astrophys. J.}\ }\textbf {\bibinfo {volume} {670}},\ \bibinfo
  {pages} {313} (\bibinfo {year} {2007})},\ \Eprint
  {http://arxiv.org/abs/0706.0516} {arXiv:0706.0516 [astro-ph]} \BibitemShut
  {NoStop}%
\bibitem [{\citenamefont {Simon}\ \emph {et~al.}(2015)\citenamefont {Simon}
  \emph {et~al.}}]{2015ApJ...808...95S}%
  \BibitemOpen
  \bibfield  {author} {\bibinfo {author} {\bibfnamefont {J.}~\bibnamefont
  {Simon}} \emph {et~al.} (\bibinfo {collaboration} {DES}),\ }\href {\doibase
  10.1088/0004-637X/808/1/95} {\bibfield  {journal} {\bibinfo  {journal}
  {Astrophys. J.}\ }\textbf {\bibinfo {volume} {808}},\ \bibinfo {eid} {95}
  (\bibinfo {year} {2015})},\ \Eprint {http://arxiv.org/abs/1504.02889}
  {arXiv:1504.02889 [astro-ph.GA]} \BibitemShut {NoStop}%
\bibitem [{\citenamefont {{Martin}}\ \emph {et~al.}(2016)\citenamefont
  {{Martin}} \emph {et~al.}}]{2016MNRAS.458L..59M}%
  \BibitemOpen
  \bibfield  {author} {\bibinfo {author} {\bibfnamefont {N.~F.}\ \bibnamefont
  {{Martin}}} \emph {et~al.},\ }\href {\doibase 10.1093/mnrasl/slw013}
  {\bibfield  {journal} {\bibinfo  {journal} {Mon. Not. Roy. Astron. Soc.}\
  }\textbf {\bibinfo {volume} {458}},\ \bibinfo {pages} {L59} (\bibinfo {year}
  {2016})},\ \Eprint {http://arxiv.org/abs/1510.01326} {arXiv:1510.01326
  [astro-ph.GA]} \BibitemShut {NoStop}%
\bibitem [{\citenamefont {{Kirby}}\ \emph {et~al.}(2017)\citenamefont
  {{Kirby}}, \citenamefont {{Cohen}}, \citenamefont {{Simon}}, \citenamefont
  {{Guhathakurta}}, \citenamefont {{Thygesen}},\ and\ \citenamefont
  {{Duggan}}}]{2017ApJ...838...83K}%
  \BibitemOpen
  \bibfield  {author} {\bibinfo {author} {\bibfnamefont {E.~N.}\ \bibnamefont
  {{Kirby}}}, \bibinfo {author} {\bibfnamefont {J.~G.}\ \bibnamefont
  {{Cohen}}}, \bibinfo {author} {\bibfnamefont {J.~D.}\ \bibnamefont
  {{Simon}}}, \bibinfo {author} {\bibfnamefont {P.}~\bibnamefont
  {{Guhathakurta}}}, \bibinfo {author} {\bibfnamefont {A.~O.}\ \bibnamefont
  {{Thygesen}}}, \ and\ \bibinfo {author} {\bibfnamefont {G.~E.}\ \bibnamefont
  {{Duggan}}},\ }\href {\doibase 10.3847/1538-4357/aa6570} {\bibfield
  {journal} {\bibinfo  {journal} {Astrophys. J.}\ }\textbf {\bibinfo {volume}
  {838}},\ \bibinfo {eid} {83} (\bibinfo {year} {2017})},\ \Eprint
  {http://arxiv.org/abs/1703.02978} {arXiv:1703.02978 [astro-ph.GA]}
  \BibitemShut {NoStop}%
\bibitem [{\citenamefont {{Kirby}}\ \emph {et~al.}(2015)\citenamefont
  {{Kirby}}, \citenamefont {{Simon}},\ and\ \citenamefont
  {{Cohen}}}]{2015ApJ...810...56K}%
  \BibitemOpen
  \bibfield  {author} {\bibinfo {author} {\bibfnamefont {E.~N.}\ \bibnamefont
  {{Kirby}}}, \bibinfo {author} {\bibfnamefont {J.~D.}\ \bibnamefont
  {{Simon}}}, \ and\ \bibinfo {author} {\bibfnamefont {J.~G.}\ \bibnamefont
  {{Cohen}}},\ }\href {\doibase 10.1088/0004-637X/810/1/56} {\bibfield
  {journal} {\bibinfo  {journal} {Astrophys. J.}\ }\textbf {\bibinfo {volume}
  {810}},\ \bibinfo {eid} {56} (\bibinfo {year} {2015})},\ \Eprint
  {http://arxiv.org/abs/1506.01021} {arXiv:1506.01021 [astro-ph.GA]}
  \BibitemShut {NoStop}%
\bibitem [{\citenamefont {Walker}\ \emph {et~al.}(2016)\citenamefont {Walker}
  \emph {et~al.}}]{2016ApJ...819...53W}%
  \BibitemOpen
  \bibfield  {author} {\bibinfo {author} {\bibfnamefont {M.~G.}\ \bibnamefont
  {Walker}} \emph {et~al.},\ }\href {\doibase 10.3847/0004-637X/819/1/53}
  {\bibfield  {journal} {\bibinfo  {journal} {Astrophys. J.}\ }\textbf
  {\bibinfo {volume} {819}},\ \bibinfo {eid} {53} (\bibinfo {year} {2016})},\
  \Eprint {http://arxiv.org/abs/1511.06296} {arXiv:1511.06296 [astro-ph.GA]}
  \BibitemShut {NoStop}%
\bibitem [{\citenamefont {Simon}\ \emph {et~al.}(2020)\citenamefont {Simon}
  \emph {et~al.}}]{2020ApJ...892..137S}%
  \BibitemOpen
  \bibfield  {author} {\bibinfo {author} {\bibfnamefont {J.}~\bibnamefont
  {Simon}} \emph {et~al.} (\bibinfo {collaboration} {DES}),\ }\href {\doibase
  10.3847/1538-4357/ab7ccb} {\bibfield  {journal} {\bibinfo  {journal}
  {Astrophys. J.}\ }\textbf {\bibinfo {volume} {892}},\ \bibinfo {eid} {137}
  (\bibinfo {year} {2020})},\ \Eprint {http://arxiv.org/abs/1911.08493}
  {arXiv:1911.08493 [astro-ph.GA]} \BibitemShut {NoStop}%
\bibitem [{\citenamefont {Koposov}\ \emph {et~al.}(2015)\citenamefont {Koposov}
  \emph {et~al.}}]{2015ApJ...811...62K}%
  \BibitemOpen
  \bibfield  {author} {\bibinfo {author} {\bibfnamefont {S.~E.}\ \bibnamefont
  {Koposov}} \emph {et~al.},\ }\href {\doibase 10.1088/0004-637X/811/1/62}
  {\bibfield  {journal} {\bibinfo  {journal} {Astrophys. J.}\ }\textbf
  {\bibinfo {volume} {811}},\ \bibinfo {eid} {62} (\bibinfo {year} {2015})},\
  \Eprint {http://arxiv.org/abs/1504.07916} {arXiv:1504.07916 [astro-ph.GA]}
  \BibitemShut {NoStop}%
\bibitem [{\citenamefont {Simon}\ \emph {et~al.}(2017)\citenamefont {Simon}
  \emph {et~al.}}]{2017ApJ...838...11S}%
  \BibitemOpen
  \bibfield  {author} {\bibinfo {author} {\bibfnamefont {J.}~\bibnamefont
  {Simon}} \emph {et~al.} (\bibinfo {collaboration} {DES}),\ }\href {\doibase
  10.3847/1538-4357/aa5be7} {\bibfield  {journal} {\bibinfo  {journal}
  {Astrophys. J.}\ }\textbf {\bibinfo {volume} {838}},\ \bibinfo {eid} {11}
  (\bibinfo {year} {2017})},\ \Eprint {http://arxiv.org/abs/1610.05301}
  {arXiv:1610.05301 [astro-ph.GA]} \BibitemShut {NoStop}%
\bibitem [{\citenamefont {Willman}\ \emph {et~al.}(2011)\citenamefont
  {Willman}, \citenamefont {Geha}, \citenamefont {Strader}, \citenamefont
  {Strigari}, \citenamefont {Simon}, \citenamefont {Kirby},\ and\ \citenamefont
  {Warres}}]{2011AJ....142..128W}%
  \BibitemOpen
  \bibfield  {author} {\bibinfo {author} {\bibfnamefont {B.}~\bibnamefont
  {Willman}}, \bibinfo {author} {\bibfnamefont {M.}~\bibnamefont {Geha}},
  \bibinfo {author} {\bibfnamefont {J.}~\bibnamefont {Strader}}, \bibinfo
  {author} {\bibfnamefont {L.~E.}\ \bibnamefont {Strigari}}, \bibinfo {author}
  {\bibfnamefont {J.~D.}\ \bibnamefont {Simon}}, \bibinfo {author}
  {\bibfnamefont {E.}~\bibnamefont {Kirby}}, \ and\ \bibinfo {author}
  {\bibfnamefont {A.}~\bibnamefont {Warres}},\ }\href {\doibase
  10.1088/0004-6256/142/4/128} {\bibfield  {journal} {\bibinfo  {journal}
  {Astron. J.}\ }\textbf {\bibinfo {volume} {142}},\ \bibinfo {eid} {128}
  (\bibinfo {year} {2011})},\ \Eprint {http://arxiv.org/abs/1007.3499}
  {arXiv:1007.3499 [astro-ph.GA]} \BibitemShut {NoStop}%
\bibitem [{\citenamefont {{Goodman}}\ and\ \citenamefont
  {{Weare}}(2010)}]{2010CAMCS...5...65G}%
  \BibitemOpen
  \bibfield  {author} {\bibinfo {author} {\bibfnamefont {J.}~\bibnamefont
  {{Goodman}}}\ and\ \bibinfo {author} {\bibfnamefont {J.}~\bibnamefont
  {{Weare}}},\ }\href {\doibase 10.2140/camcos.2010.5.65} {\bibfield  {journal}
  {\bibinfo  {journal} {Communications in Applied Mathematics and Computational
  Science}\ }\textbf {\bibinfo {volume} {5}},\ \bibinfo {pages} {65} (\bibinfo
  {year} {2010})}\BibitemShut {NoStop}%
\bibitem [{\citenamefont {de~Naray}\ \emph {et~al.}(2008)\citenamefont
  {de~Naray}, \citenamefont {McGaugh},\ and\ \citenamefont
  {de~Blok}}]{de_Naray_2008}%
  \BibitemOpen
  \bibfield  {author} {\bibinfo {author} {\bibfnamefont {R.~K.}\ \bibnamefont
  {de~Naray}}, \bibinfo {author} {\bibfnamefont {S.~S.}\ \bibnamefont
  {McGaugh}}, \ and\ \bibinfo {author} {\bibfnamefont {W.~J.~G.}\ \bibnamefont
  {de~Blok}},\ }\href {\doibase 10.1086/527543} {\bibfield  {journal} {\bibinfo
   {journal} {The Astrophysical Journal}\ }\textbf {\bibinfo {volume} {676}},\
  \bibinfo {pages} {920–943} (\bibinfo {year} {2008})}\BibitemShut {NoStop}%
\bibitem [{\citenamefont {Newman}\ \emph {et~al.}(2013)\citenamefont {Newman},
  \citenamefont {Treu}, \citenamefont {Ellis},\ and\ \citenamefont
  {Sand}}]{Newman_2013}%
  \BibitemOpen
  \bibfield  {author} {\bibinfo {author} {\bibfnamefont {A.~B.}\ \bibnamefont
  {Newman}}, \bibinfo {author} {\bibfnamefont {T.}~\bibnamefont {Treu}},
  \bibinfo {author} {\bibfnamefont {R.~S.}\ \bibnamefont {Ellis}}, \ and\
  \bibinfo {author} {\bibfnamefont {D.~J.}\ \bibnamefont {Sand}},\ }\href
  {\doibase 10.1088/0004-637x/765/1/25} {\bibfield  {journal} {\bibinfo
  {journal} {The Astrophysical Journal}\ }\textbf {\bibinfo {volume} {765}},\
  \bibinfo {pages} {25} (\bibinfo {year} {2013})}\BibitemShut {NoStop}%
\bibitem [{Note3()}]{Note3}%
  \BibitemOpen
  \bibinfo {note} {For a model which predicts the SIDM cross-section with an
  intricate velocity dependence, see e.g. Ref.\cite {Chu:2018fzy}.}\BibitemShut
  {Stop}%
\bibitem [{\citenamefont {Correa}(2020)}]{Correa:2020qam}%
  \BibitemOpen
  \bibfield  {author} {\bibinfo {author} {\bibfnamefont {C.~A.}\ \bibnamefont
  {Correa}},\ }\href@noop {} {\  (\bibinfo {year} {2020})},\ \Eprint
  {http://arxiv.org/abs/2007.02958} {arXiv:2007.02958 [astro-ph.GA]}
  \BibitemShut {NoStop}%
\bibitem [{\citenamefont {Balberg}\ \emph {et~al.}(2002)\citenamefont
  {Balberg}, \citenamefont {Shapiro},\ and\ \citenamefont
  {Inagaki}}]{Balberg:2002ue}%
  \BibitemOpen
  \bibfield  {author} {\bibinfo {author} {\bibfnamefont {S.}~\bibnamefont
  {Balberg}}, \bibinfo {author} {\bibfnamefont {S.~L.}\ \bibnamefont
  {Shapiro}}, \ and\ \bibinfo {author} {\bibfnamefont {S.}~\bibnamefont
  {Inagaki}},\ }\href {\doibase 10.1086/339038} {\bibfield  {journal} {\bibinfo
   {journal} {Astrophys. J.}\ }\textbf {\bibinfo {volume} {568}},\ \bibinfo
  {pages} {475} (\bibinfo {year} {2002})},\ \Eprint
  {http://arxiv.org/abs/astro-ph/0110561} {arXiv:astro-ph/0110561} \BibitemShut
  {NoStop}%
\bibitem [{\citenamefont {Nishikawa}\ \emph {et~al.}(2020)\citenamefont
  {Nishikawa}, \citenamefont {Boddy},\ and\ \citenamefont
  {Kaplinghat}}]{Nishikawa:2019lsc}%
  \BibitemOpen
  \bibfield  {author} {\bibinfo {author} {\bibfnamefont {H.}~\bibnamefont
  {Nishikawa}}, \bibinfo {author} {\bibfnamefont {K.~K.}\ \bibnamefont
  {Boddy}}, \ and\ \bibinfo {author} {\bibfnamefont {M.}~\bibnamefont
  {Kaplinghat}},\ }\href {\doibase 10.1103/PhysRevD.101.063009} {\bibfield
  {journal} {\bibinfo  {journal} {Phys. Rev. D}\ }\textbf {\bibinfo {volume}
  {101}},\ \bibinfo {pages} {063009} (\bibinfo {year} {2020})},\ \Eprint
  {http://arxiv.org/abs/1901.00499} {arXiv:1901.00499 [astro-ph.GA]}
  \BibitemShut {NoStop}%
\bibitem [{\citenamefont {Kummer}\ \emph {et~al.}(2019)\citenamefont {Kummer},
  \citenamefont {Brüggen}, \citenamefont {Dolag}, \citenamefont {Kahlhoefer},\
  and\ \citenamefont {Schmidt-Hoberg}}]{Kummer:2019yrb}%
  \BibitemOpen
  \bibfield  {author} {\bibinfo {author} {\bibfnamefont {J.}~\bibnamefont
  {Kummer}}, \bibinfo {author} {\bibfnamefont {M.}~\bibnamefont {Brüggen}},
  \bibinfo {author} {\bibfnamefont {K.}~\bibnamefont {Dolag}}, \bibinfo
  {author} {\bibfnamefont {F.}~\bibnamefont {Kahlhoefer}}, \ and\ \bibinfo
  {author} {\bibfnamefont {K.}~\bibnamefont {Schmidt-Hoberg}},\ }\href
  {\doibase 10.1093/mnras/stz1261} {\bibfield  {journal} {\bibinfo  {journal}
  {Mon. Not. Roy. Astron. Soc.}\ }\textbf {\bibinfo {volume} {487}},\ \bibinfo
  {pages} {354} (\bibinfo {year} {2019})},\ \Eprint
  {http://arxiv.org/abs/1902.02330} {arXiv:1902.02330 [astro-ph.CO]}
  \BibitemShut {NoStop}%
\bibitem [{\citenamefont {Robles}\ \emph {et~al.}(2019)\citenamefont {Robles},
  \citenamefont {Kelley}, \citenamefont {Bullock},\ and\ \citenamefont
  {Kaplinghat}}]{Robles:2019mfq}%
  \BibitemOpen
  \bibfield  {author} {\bibinfo {author} {\bibfnamefont {V.~H.}\ \bibnamefont
  {Robles}}, \bibinfo {author} {\bibfnamefont {T.}~\bibnamefont {Kelley}},
  \bibinfo {author} {\bibfnamefont {J.~S.}\ \bibnamefont {Bullock}}, \ and\
  \bibinfo {author} {\bibfnamefont {M.}~\bibnamefont {Kaplinghat}},\ }\href
  {\doibase 10.1093/mnras/stz2345} {\bibfield  {journal} {\bibinfo  {journal}
  {Mon. Not. Roy. Astron. Soc.}\ }\textbf {\bibinfo {volume} {490}},\ \bibinfo
  {pages} {2117} (\bibinfo {year} {2019})},\ \Eprint
  {http://arxiv.org/abs/1903.01469} {arXiv:1903.01469 [astro-ph.GA]}
  \BibitemShut {NoStop}%
\bibitem [{\citenamefont {Sameie}\ \emph {et~al.}(2020)\citenamefont {Sameie},
  \citenamefont {Yu}, \citenamefont {Sales}, \citenamefont {Vogelsberger},\
  and\ \citenamefont {Zavala}}]{Sameie:2019zfo}%
  \BibitemOpen
  \bibfield  {author} {\bibinfo {author} {\bibfnamefont {O.}~\bibnamefont
  {Sameie}}, \bibinfo {author} {\bibfnamefont {H.-B.}\ \bibnamefont {Yu}},
  \bibinfo {author} {\bibfnamefont {L.~V.}\ \bibnamefont {Sales}}, \bibinfo
  {author} {\bibfnamefont {M.}~\bibnamefont {Vogelsberger}}, \ and\ \bibinfo
  {author} {\bibfnamefont {J.}~\bibnamefont {Zavala}},\ }\href {\doibase
  10.1103/PhysRevLett.124.141102} {\bibfield  {journal} {\bibinfo  {journal}
  {Phys. Rev. Lett.}\ }\textbf {\bibinfo {volume} {124}},\ \bibinfo {pages}
  {141102} (\bibinfo {year} {2020})},\ \Eprint
  {http://arxiv.org/abs/1904.07872} {arXiv:1904.07872 [astro-ph.GA]}
  \BibitemShut {NoStop}%
\bibitem [{\citenamefont {Kahlhoefer}\ \emph {et~al.}(2019)\citenamefont
  {Kahlhoefer}, \citenamefont {Kaplinghat}, \citenamefont {Slatyer},\ and\
  \citenamefont {Wu}}]{Kahlhoefer:2019oyt}%
  \BibitemOpen
  \bibfield  {author} {\bibinfo {author} {\bibfnamefont {F.}~\bibnamefont
  {Kahlhoefer}}, \bibinfo {author} {\bibfnamefont {M.}~\bibnamefont
  {Kaplinghat}}, \bibinfo {author} {\bibfnamefont {T.~R.}\ \bibnamefont
  {Slatyer}}, \ and\ \bibinfo {author} {\bibfnamefont {C.-L.}\ \bibnamefont
  {Wu}},\ }\href {\doibase 10.1088/1475-7516/2019/12/010} {\bibfield  {journal}
  {\bibinfo  {journal} {JCAP}\ }\textbf {\bibinfo {volume} {12}},\ \bibinfo
  {pages} {010} (\bibinfo {year} {2019})},\ \Eprint
  {http://arxiv.org/abs/1904.10539} {arXiv:1904.10539 [astro-ph.GA]}
  \BibitemShut {NoStop}%
\bibitem [{\citenamefont {Kaplinghat}\ \emph {et~al.}(2014)\citenamefont
  {Kaplinghat}, \citenamefont {Keeley}, \citenamefont {Linden},\ and\
  \citenamefont {Yu}}]{Kaplinghat_2014}%
  \BibitemOpen
  \bibfield  {author} {\bibinfo {author} {\bibfnamefont {M.}~\bibnamefont
  {Kaplinghat}}, \bibinfo {author} {\bibfnamefont {R.~E.}\ \bibnamefont
  {Keeley}}, \bibinfo {author} {\bibfnamefont {T.}~\bibnamefont {Linden}}, \
  and\ \bibinfo {author} {\bibfnamefont {H.-B.}\ \bibnamefont {Yu}},\ }\href
  {\doibase 10.1103/physrevlett.113.021302} {\bibfield  {journal} {\bibinfo
  {journal} {Physical Review Letters}\ }\textbf {\bibinfo {volume} {113}}
  (\bibinfo {year} {2014}),\ 10.1103/physrevlett.113.021302}\BibitemShut
  {NoStop}%
\bibitem [{\citenamefont {Vogelsberger}\ \emph
  {et~al.}(2014{\natexlab{c}})\citenamefont {Vogelsberger}, \citenamefont
  {Zavala}, \citenamefont {Simpson},\ and\ \citenamefont
  {Jenkins}}]{Vogelsberger_2014}%
  \BibitemOpen
  \bibfield  {author} {\bibinfo {author} {\bibfnamefont {M.}~\bibnamefont
  {Vogelsberger}}, \bibinfo {author} {\bibfnamefont {J.}~\bibnamefont
  {Zavala}}, \bibinfo {author} {\bibfnamefont {C.}~\bibnamefont {Simpson}}, \
  and\ \bibinfo {author} {\bibfnamefont {A.}~\bibnamefont {Jenkins}},\ }\href
  {\doibase 10.1093/mnras/stu1713} {\bibfield  {journal} {\bibinfo  {journal}
  {Monthly Notices of the Royal Astronomical Society}\ }\textbf {\bibinfo
  {volume} {444}},\ \bibinfo {pages} {3684–3698} (\bibinfo {year}
  {2014}{\natexlab{c}})}\BibitemShut {NoStop}%
\bibitem [{\citenamefont {Kamada}\ \emph {et~al.}(2017)\citenamefont {Kamada},
  \citenamefont {Kaplinghat}, \citenamefont {Pace},\ and\ \citenamefont
  {Yu}}]{Kamada_2017}%
  \BibitemOpen
  \bibfield  {author} {\bibinfo {author} {\bibfnamefont {A.}~\bibnamefont
  {Kamada}}, \bibinfo {author} {\bibfnamefont {M.}~\bibnamefont {Kaplinghat}},
  \bibinfo {author} {\bibfnamefont {A.~B.}\ \bibnamefont {Pace}}, \ and\
  \bibinfo {author} {\bibfnamefont {H.-B.}\ \bibnamefont {Yu}},\ }\href
  {\doibase 10.1103/physrevlett.119.111102} {\bibfield  {journal} {\bibinfo
  {journal} {Physical Review Letters}\ }\textbf {\bibinfo {volume} {119}}
  (\bibinfo {year} {2017}),\ 10.1103/physrevlett.119.111102}\BibitemShut
  {NoStop}%
\bibitem [{\citenamefont {Creasey}\ \emph {et~al.}(2017)\citenamefont
  {Creasey}, \citenamefont {Sameie}, \citenamefont {Sales}, \citenamefont {Yu},
  \citenamefont {Vogelsberger},\ and\ \citenamefont {Zavala}}]{Creasey_2017}%
  \BibitemOpen
  \bibfield  {author} {\bibinfo {author} {\bibfnamefont {P.}~\bibnamefont
  {Creasey}}, \bibinfo {author} {\bibfnamefont {O.}~\bibnamefont {Sameie}},
  \bibinfo {author} {\bibfnamefont {L.~V.}\ \bibnamefont {Sales}}, \bibinfo
  {author} {\bibfnamefont {H.-B.}\ \bibnamefont {Yu}}, \bibinfo {author}
  {\bibfnamefont {M.}~\bibnamefont {Vogelsberger}}, \ and\ \bibinfo {author}
  {\bibfnamefont {J.}~\bibnamefont {Zavala}},\ }\href {\doibase
  10.1093/mnras/stx522} {\bibfield  {journal} {\bibinfo  {journal} {Monthly
  Notices of the Royal Astronomical Society}\ }\textbf {\bibinfo {volume}
  {468}},\ \bibinfo {pages} {2283–2295} (\bibinfo {year} {2017})}\BibitemShut
  {NoStop}%
\bibitem [{\citenamefont {Elbert}\ \emph {et~al.}(2018)\citenamefont {Elbert},
  \citenamefont {Bullock}, \citenamefont {Kaplinghat}, \citenamefont
  {Garrison-Kimmel}, \citenamefont {Graus},\ and\ \citenamefont
  {Rocha}}]{Elbert_2018}%
  \BibitemOpen
  \bibfield  {author} {\bibinfo {author} {\bibfnamefont {O.~D.}\ \bibnamefont
  {Elbert}}, \bibinfo {author} {\bibfnamefont {J.~S.}\ \bibnamefont {Bullock}},
  \bibinfo {author} {\bibfnamefont {M.}~\bibnamefont {Kaplinghat}}, \bibinfo
  {author} {\bibfnamefont {S.}~\bibnamefont {Garrison-Kimmel}}, \bibinfo
  {author} {\bibfnamefont {A.~S.}\ \bibnamefont {Graus}}, \ and\ \bibinfo
  {author} {\bibfnamefont {M.}~\bibnamefont {Rocha}},\ }\href {\doibase
  10.3847/1538-4357/aa9710} {\bibfield  {journal} {\bibinfo  {journal} {The
  Astrophysical Journal}\ }\textbf {\bibinfo {volume} {853}},\ \bibinfo {pages}
  {109} (\bibinfo {year} {2018})}\BibitemShut {NoStop}%
\bibitem [{\citenamefont {Robertson}\ \emph {et~al.}(2018)\citenamefont
  {Robertson}, \citenamefont {Massey}, \citenamefont {Eke}, \citenamefont
  {Tulin}, \citenamefont {Yu}, \citenamefont {Bahé}, \citenamefont {Barnes},
  \citenamefont {Bower}, \citenamefont {Crain}, \citenamefont {Dalla~Vecchia},\
  and\ \citenamefont {et~al.}}]{Robertson_2018}%
  \BibitemOpen
  \bibfield  {author} {\bibinfo {author} {\bibfnamefont {A.}~\bibnamefont
  {Robertson}}, \bibinfo {author} {\bibfnamefont {R.}~\bibnamefont {Massey}},
  \bibinfo {author} {\bibfnamefont {V.}~\bibnamefont {Eke}}, \bibinfo {author}
  {\bibfnamefont {S.}~\bibnamefont {Tulin}}, \bibinfo {author} {\bibfnamefont
  {H.-B.}\ \bibnamefont {Yu}}, \bibinfo {author} {\bibfnamefont
  {Y.}~\bibnamefont {Bahé}}, \bibinfo {author} {\bibfnamefont {D.~J.}\
  \bibnamefont {Barnes}}, \bibinfo {author} {\bibfnamefont {R.~G.}\
  \bibnamefont {Bower}}, \bibinfo {author} {\bibfnamefont {R.~A.}\ \bibnamefont
  {Crain}}, \bibinfo {author} {\bibfnamefont {C.}~\bibnamefont
  {Dalla~Vecchia}}, \ and\ \bibinfo {author} {\bibnamefont {et~al.}},\ }\href
  {\doibase 10.1093/mnrasl/sly024} {\bibfield  {journal} {\bibinfo  {journal}
  {Monthly Notices of the Royal Astronomical Society: Letters}\ }\textbf
  {\bibinfo {volume} {476}},\ \bibinfo {pages} {L20–L24} (\bibinfo {year}
  {2018})}\BibitemShut {NoStop}%
\bibitem [{\citenamefont {Ren}\ \emph {et~al.}(2019)\citenamefont {Ren},
  \citenamefont {Kwa}, \citenamefont {Kaplinghat},\ and\ \citenamefont
  {Yu}}]{Ren_2019}%
  \BibitemOpen
  \bibfield  {author} {\bibinfo {author} {\bibfnamefont {T.}~\bibnamefont
  {Ren}}, \bibinfo {author} {\bibfnamefont {A.}~\bibnamefont {Kwa}}, \bibinfo
  {author} {\bibfnamefont {M.}~\bibnamefont {Kaplinghat}}, \ and\ \bibinfo
  {author} {\bibfnamefont {H.-B.}\ \bibnamefont {Yu}},\ }\href {\doibase
  10.1103/physrevx.9.031020} {\bibfield  {journal} {\bibinfo  {journal}
  {Physical Review X}\ }\textbf {\bibinfo {volume} {9}} (\bibinfo {year}
  {2019}),\ 10.1103/physrevx.9.031020}\BibitemShut {NoStop}%
\bibitem [{\citenamefont {Oman}\ \emph {et~al.}(2015)\citenamefont {Oman},
  \citenamefont {Navarro}, \citenamefont {Fattahi}, \citenamefont {Frenk},
  \citenamefont {Sawala}, \citenamefont {White}, \citenamefont {Bower},
  \citenamefont {Crain}, \citenamefont {Furlong}, \citenamefont {Schaller},\
  and\ \citenamefont {et~al.}}]{Oman_2015}%
  \BibitemOpen
  \bibfield  {author} {\bibinfo {author} {\bibfnamefont {K.~A.}\ \bibnamefont
  {Oman}}, \bibinfo {author} {\bibfnamefont {J.~F.}\ \bibnamefont {Navarro}},
  \bibinfo {author} {\bibfnamefont {A.}~\bibnamefont {Fattahi}}, \bibinfo
  {author} {\bibfnamefont {C.~S.}\ \bibnamefont {Frenk}}, \bibinfo {author}
  {\bibfnamefont {T.}~\bibnamefont {Sawala}}, \bibinfo {author} {\bibfnamefont
  {S.~D.~M.}\ \bibnamefont {White}}, \bibinfo {author} {\bibfnamefont
  {R.}~\bibnamefont {Bower}}, \bibinfo {author} {\bibfnamefont {R.~A.}\
  \bibnamefont {Crain}}, \bibinfo {author} {\bibfnamefont {M.}~\bibnamefont
  {Furlong}}, \bibinfo {author} {\bibfnamefont {M.}~\bibnamefont {Schaller}}, \
  and\ \bibinfo {author} {\bibnamefont {et~al.}},\ }\href {\doibase
  10.1093/mnras/stv1504} {\bibfield  {journal} {\bibinfo  {journal} {Monthly
  Notices of the Royal Astronomical Society}\ }\textbf {\bibinfo {volume}
  {452}},\ \bibinfo {pages} {3650–3665} (\bibinfo {year} {2015})}\BibitemShut
  {NoStop}%
\bibitem [{\citenamefont {Oman}\ \emph {et~al.}(2016)\citenamefont {Oman},
  \citenamefont {Navarro}, \citenamefont {Sales}, \citenamefont {Fattahi},
  \citenamefont {Frenk}, \citenamefont {Sawala}, \citenamefont {Schaller},\
  and\ \citenamefont {White}}]{Oman_2016}%
  \BibitemOpen
  \bibfield  {author} {\bibinfo {author} {\bibfnamefont {K.~A.}\ \bibnamefont
  {Oman}}, \bibinfo {author} {\bibfnamefont {J.~F.}\ \bibnamefont {Navarro}},
  \bibinfo {author} {\bibfnamefont {L.~V.}\ \bibnamefont {Sales}}, \bibinfo
  {author} {\bibfnamefont {A.}~\bibnamefont {Fattahi}}, \bibinfo {author}
  {\bibfnamefont {C.~S.}\ \bibnamefont {Frenk}}, \bibinfo {author}
  {\bibfnamefont {T.}~\bibnamefont {Sawala}}, \bibinfo {author} {\bibfnamefont
  {M.}~\bibnamefont {Schaller}}, \ and\ \bibinfo {author} {\bibfnamefont
  {S.~D.~M.}\ \bibnamefont {White}},\ }\href {\doibase 10.1093/mnras/stw1251}
  {\bibfield  {journal} {\bibinfo  {journal} {Monthly Notices of the Royal
  Astronomical Society}\ }\textbf {\bibinfo {volume} {460}},\ \bibinfo {pages}
  {3610–3623} (\bibinfo {year} {2016})}\BibitemShut {NoStop}%
\bibitem [{\citenamefont {Chu}\ \emph {et~al.}(2019)\citenamefont {Chu},
  \citenamefont {Garcia-Cely},\ and\ \citenamefont {Murayama}}]{Chu:2018fzy}%
  \BibitemOpen
  \bibfield  {author} {\bibinfo {author} {\bibfnamefont {X.}~\bibnamefont
  {Chu}}, \bibinfo {author} {\bibfnamefont {C.}~\bibnamefont {Garcia-Cely}}, \
  and\ \bibinfo {author} {\bibfnamefont {H.}~\bibnamefont {Murayama}},\ }\href
  {\doibase 10.1103/PhysRevLett.122.071103} {\bibfield  {journal} {\bibinfo
  {journal} {Phys. Rev. Lett.}\ }\textbf {\bibinfo {volume} {122}},\ \bibinfo
  {pages} {071103} (\bibinfo {year} {2019})},\ \Eprint
  {http://arxiv.org/abs/1810.04709} {arXiv:1810.04709 [hep-ph]} \BibitemShut
  {NoStop}%
\end{thebibliography}%
%%%%%%%%%%%%%%%%%%%%%%%%%%%%%%%%%%%%%%%%%%%%

\end{document}